\documentclass[12pt,draft]{IEEEtran} 
\usepackage{amsmath}

\def\BibTeX{{\rm B\kern-.05em{\sc i\kern-.025em b}\kern-.08em
    T\kern-.1667em\lower.7ex\hbox{E}\kern-.125emX}}
\begin{document}

\title{\Large\bf Interaction in Quantum Communication}

\author{
     Hartmut Klauck, Ashwin Nayak, Amnon Ta-Shma and David Zuckerman
\thanks{Hartmut is with the Department of Computer Science and
Mathematics, University of Frankfurt, Robert Mayer Strasse 11-15,
60054 Frankfurt am Main, Germany.
His research is supported by DFG grant KL 1470/1.
E-mail:   klauck@thi.informatik.uni-frankfurt.de.
Most of this work was done while Hartmut was with the University of 
Frankfurt, and later with CWI, supported by the EU 5th framework 
program QAIP IST-1999-11234 and by NWO grant 612.055.001.
  Ashwin is with Department of Combinatorics and Optimization, and
Institute for Quantum Computing, University of Waterloo, 200
University Ave.\ W., Waterloo, ON N2L 3G1, Canada, E-mail:
anayak@math.uwaterloo.ca. He is also Associate Member, Perimeter
Institute for Theoretical Physics, Canada.
Ashwin's research is supported in part by NSERC, CIAR, MITACS, CFI, and
OIT (Canada).
Parts of this work were done while Ashwin was at University of
California, Berkeley, DIMACS Center and AT{\&}T Labs, and California
Institute of Technology.
     Amnon is with the Dept. of Computer Science,
     Tel-Aviv University,
     Israel 69978,
     E-mail: amnon@post.tau.ac.il.
This research was supported in part by Grant No 2004390 from the United
States-Israel Binational Science Foundation (BSF), Jerusalem, Israel.
A part of this work was done while Amnon was at
 the University of California at Berkeley, and
 supported in part by a David and Lucile
 Packard Fellowship for Science and Engineering and
 NSF NYI Grant CCR-9457799.
     David is with the Dept. of Computer Science,
     University of Texas,
     Austin, TX  78712,
     E-mail: diz@cs.utexas.edu. This work was done while David was on leave at
 the University of California at Berkeley.
 Supported in part by a David and Lucile
 Packard Fellowship for Science and Engineering,
 NSF Grant CCR-9912428,
 NSF NYI Grant CCR-9457799,
 and an Alfred P.\ Sloan Research Fellowship.}
}

\markboth{}
{Klauck, Nayak, Ta-Shma and Zuckerman: Round hierarchy in Quantum communication  } 

\maketitle

\newtheorem{theorem}{Theorem}[section]
\newtheorem{lemma}[theorem]{Lemma}
\newtheorem{fact}[theorem]{Fact}
\newtheorem{proposition}[theorem]{Proposition}
\newtheorem{claim}[theorem]{Claim}
\newtheorem{corollary}[theorem]{Corollary}
\newtheorem{definition}{Definition}[section]
\newtheorem{observation}[theorem]{Observation}

\def\square{\rule{2mm}{2mm}}
\newenvironment{remark}{{\noindent\bf Remark:  }}{\qquad\square}
\newenvironment{example}{{\noindent\bf Example:  }}{\qquad\square}

\newcommand{\tensor}{\otimes}
\newcommand{\xor}{\oplus}
\newcommand{\union}{\cup}
\newcommand{\intersect}{\cap}
\newcommand\meet\wedge
\newcommand{\up}{\uparrow}
\newcommand{\down}{\downarrow}

\newcommand{\reals}{{\hbox{\sf I\kern-.14em\hbox{R}}}}
\newcommand{\trace}{{\rm Tr}}
\newcommand{\prob}{{\rm Prob}}
\newcommand{\expct}[1]{{\mathrm E}\left[#1\right]}
\newcommand{\size}[1]{\left|#1\right|}
\newcommand{\ceil}[1]{\left\lceil#1\right\rceil}
\newcommand{\floor}[1]{\left\lfloor#1\right\rfloor}
\newcommand{\ket}[1]{\left|#1\right\rangle}
\newcommand{\bra}[1]{\left\langle #1\right|}
\newcommand{\braket}[2]{\left\langle #1\!\mid\! #2\right\rangle}
\newcommand{\ketbra}[2]{\ket{#1}\!\bra{#2}}
\newcommand{\norm}[1]{\left\|\,#1\,\right\|}
\newcommand{\trnorm}[1]{\norm{#1}_{\rm t}}
\newcommand{\hel}[1]{h\left(#1\right)}
\newcommand{\hels}[1]{h^2\left(#1\right)}
\newcommand{\set}[1]{\left\{#1\right\}}
\newcommand{\st}{{\; | \;}}
\newcommand{\eqdef}{\stackrel{\rm def}{=}}
\newcommand{\eqq}{\stackrel{\rm ?}{=}}

\newcommand{\remove}[1]{}
\newcommand{\eq}{\;=\;}
\newcommand{\qu}{\quad}

\newcommand{\eqcc}{\tilde{Q}}
\newcommand{\co}{{\cal O}}
\newcommand{\ch}{{\cal H}}
\newcommand{\disj}{{\rm DISJ}}
\newcommand{\oh}{{\mathcal O}}
\newcommand{\aitch}{{\mathcal H}}
\newcommand{\kay}{{\mathcal K}}
\newcommand{\bsigma}{{\mathbf \sigma}}
\newcommand{\brho}{{\mathbf \rho}}
\newcommand{\ebar}{{\bar{\epsilon}}}
\newcommand{\hx}{{\hat{x}}}
\newcommand{\hy}{{\hat{y}}}
\newcommand{\trn}[1]{\trnorm{#1}}
\newcommand{\la}{\langle}
\newcommand{\ra}{\rangle}
\newcommand{\density}[1]{\ketbra{#1}{#1}}
\newcommand{\ol}[1]{\overline{#1}}
\newcommand{\X}{{\mathcal X}}
\newcommand{\Y}{{\mathcal Y}}
\newcommand{\cp}{{\mathcal P}}
\newcommand{\E}{{\mathbf E}}
\newcommand{\F}{{\mathcal{F}}}

\newcommand{\etal}{{\it et al.\/}}

\begin{abstract}
In some scenarios there are ways of conveying information with
many fewer, even exponentially fewer, qubits than possible
classically~\cite{bcw,ASTVW03,R99}. Moreover, some of these
methods have a very simple structure---they involve only few
message exchanges between the communicating parties. It is
therefore natural to ask whether {\em every\/} classical protocol
may be transformed to a ``simpler'' quantum protocol---one that
has similar efficiency, but uses fewer message exchanges.

We show that for any constant~$k$, there is a problem such that
its~$k+1$ message classical communication complexity is
exponentially smaller than its~$k$ message quantum communication
complexity. This, in particular, proves a round hierarchy theorem
for quantum communication complexity, and implies, via a simple
reduction, an~$\Omega(N^{1/k})$ lower bound for~$k$ message
quantum protocols for Set Disjointness for constant~$k$.

Enroute, we prove information-theoretic lemmas, and define a
related measure of {\it correlation}, the {\em informational
distance\/}, that we believe may be of significance in other
contexts as well.
\end{abstract}

\section{Introduction}
\label{sec:intro}

A recurring theme in quantum information processing has been the
idea of exploiting the exponential resources afforded by quantum
states to encode information in very non-obvious ways. One
representative result of this kind is due to Ambainis, Schulman,
Ta-Shma, Vazirani, and Wigderson~\cite{ASTVW03}. They show that
two players can deal a random set of~$\sqrt{N}$ cards each, from a
pack of~$N$ cards, by the exchange of~$O(\log N)$ quantum bits
between them. Another example is given by Raz~\cite{R99} who shows
that a natural geometric promise problem that has an efficient
quantum protocol, is hard to solve via classical communication.
Both are examples of problems for which exponentially fewer
quantum bits are required to accomplish a communication task, as
compared to classical bits. A third example is the~$O(\sqrt{N}\log
N)$ qubit protocol for Set Disjointness due to Buhrman, Cleve, and
Wigderson~\cite{bcw}, which represents quadratic savings in the
communication cost over classical protocols.

The protocols presented by Ambainis \etal{}~\cite{ASTVW03} and
Raz~\cite{R99} share the feature that they require minimal {\em
interaction\/} between the communicating players. For example, in the
protocol of Ambainis \etal{}~\cite{ASTVW03} one player prepares a set
of qubits in a certain state and sends half of the qubits across as
the message, after which both players measure their qubits to obtain
the result. In contrast, the protocol of Buhrman, Cleve and
Wigderson~\cite{bcw} for checking set disjointness (\disj) requires
$\Omega(\sqrt{N})$ messages. This raises a natural question: Can we
exploit the features of quantum communication and always reduce
interaction while maintaining the same communication cost? In
particular, are there efficient quantum protocols for \disj\ that
require only a few messages?

Kitaev and Watrous~\cite{KitaevW00} show that every efficient
quantum interactive proof can be transformed into a protocol with
only three messages of similar total length. This suggests that it
might be possible to reduce interaction in other protocols as
well. In this paper we show that for any constant~$k$, there is a
problem such that its~$k+1$ message classical communication
complexity is exponentially smaller than its~$k$ message quantum
communication complexity, thus answering the above question in the
negative. This, in particular, proves a round hierarchy theorem
for quantum communication complexity, and implies, via a simple
reduction, polynomial lower bounds for constant round quantum
protocols for Set Disjointness.

\subsection*{Our Separation Results}

The role of interaction in {\em classical} communication is
well-studied, especially in the context of the Pointer Jumping
function~\cite{PS82,DGS87,NW93,Kl98,PRV99}. Our first result is
for a subproblem~$S_k$ of Pointer Jumping that is singled out in
Miltersen  \etal{}~\cite{MNSW95} (see Section~\ref{sec:problem}
for a formal definition of~$S_k$). We show:

\begin{theorem}
\label{thm:main} For any constant~$k$, there is a problem
$S_{k+1}$ such that any quantum protocol with only~$k$ messages
and constant probability of error requires~$\Omega(N^{1/(k+1)})$
communication qubits, whereas it can be solved with~$k+1$ messages
by a deterministic protocol with~$O(\log N)$ bits.
\end{theorem}

A more precise version of this theorem is given in
Section~\ref{sec:non-const} and implies a round hierarchy even
when the number of messages~$k$ grows as a function of input
size~$N$, up to~$k = \Theta(\log N/\log\log N)$. Our analysis
of~$S_k$ follows the same intuition as that behind the result of
Miltersen \etal{}~\cite{MNSW95},
but relies on entirely new ideas from quantum information theory.
The resulting lower bound is optimal for a constant number of
rounds.

Next, we study the Pointer Jumping function itself. Let~$f_k$
denote the Pointer Jumping function with path length~$k+1$ on
graphs with~$2n$ vertices, as defined in
Section~\ref{sec:pointer-jumping}. The input length for the
Pointer Jumping function~$f_k$ is~$N = 2n\log n$, independent
of~$k$, whereas the input length for $S_k$ is exponential in~$k$.
The function~$f_k$ is thus usually more appropriate for studying
the effect of rounds on communication when~$k$ grows rapidly as a
function of the input length.

We first show an improved upper bound on the classical complexity
of Pointer Jumping, further closing the gap between the known
classical upper and lower bounds. We then turn into proving a
quantum lower bound. We prove:

\begin{theorem}
\label{thm:PJ-result}
For any constant~$k$,
there is a classical deterministic protocol with~$k$ message
exchanges, that computes~$f_k$ with $O(\log n)$ bits of communication,
while any~$k-1$ round quantum protocol with constant error for $f_k$
needs $\Omega(n)$ qubits communication.
\end{theorem}

The lower bound of Theorem~\ref{thm:PJ-result} decays
exponentially in~$k$, and leads only to separation results for~$k
= O(\log N)$. We believe it is possible to improve this dependence
on~$k$, but leave it as an open problem. Note that in the
preliminary version of this paper~\cite{KNTZ01} this decay was
even doubly exponential, and the improvement here is obtained by
using a quantum version of the Hellinger distance.


Our lower bounds for~$S_k$ and Pointer Jumping also have implications
for Set Disjointness. The problem of determining the quantum
communication complexity of \disj\ has inspired much research in the
last few years, yet the best known lower bound prior to this work
was~$\Omega(\log n)$~\cite{ASTVW03,BW01}. We mentioned earlier the
protocol of Buhrman \etal{}~\cite{bcw} which solves \disj\ with
$O(\sqrt{N} \log N)$ qubits and $\Omega(\sqrt{N})$ messages.  Buhrman
and de Wolf~\cite{BW01} observed (based on a lower bound for random
access codes~\cite{Nayak99,AmbainisNTV02})
that any one message quantum protocol for \disj\
has linear communication complexity. We describe a simple reduction
from Pointer Jumping in a bounded number of rounds to \disj\ and
prove:

\begin{corollary}
\label{thm-disj}
For any constant~$k$,
the communication complexity of any~$k$-message quantum protocol for Set
Disjointness is~$\Omega(N^{1/k})$.
\end{corollary}

A model of quantum communication complexity that has also been
studied in the literature is that of communication with prior
entanglement (see, e.g., Refs.~\cite{CDNT98,BW01}). In this model, the
communicating parties may hold an arbitrary input-independent
entangled state in the beginning of a protocol. One can use
superdense coding~\cite{BW92} to transmit~$n$ classical bits of
information using only~$\lceil n/2\rceil$ qubits when entanglement
is allowed. The players may also use measurements on EPR-pairs to
create a shared classical random key. While the first idea often
decreases the communication complexity by a factor of two, the
second sometimes saves~$\log n$ bits of communication. It is
unknown if shared entanglement may sometimes decrease the
communication more than that. Currently no general methods for
proving super-logarithmic lower bounds on the quantum
communication complexity with prior entanglement and unrestricted
interaction are known. Our results all hold in this model as well.

Our interest in the role of interaction in quantum communication
also springs from the need to better understand the ways in which
we can access and manipulate information encoded in quantum
states. We develop information-theoretic techniques that expose
some of the limitations of quantum communication. We believe our
information-theoretic results are of independent interest.

The paper is organized as follows. In Section
\ref{sec:info-background} we give some background on classical and
quantum information theory. We recommend Preskill's lecture
notes~\cite{P98} or Nielsen and Chuang's book~\cite{NC00} as thorough
introductions into the field. In Section
\ref{sec:information-theory-new} we present new lower bounds on the
quantum relative entropy function (Section \ref{sec:relative-trace})
and introduce the informational distance (Section
\ref{sec:informational-distance}). In Section \ref{sec:cc-model} we
explain the communication complexity model, followed by Section
\ref{sec:Sk} where we prove our separation results and the reduction
to Set Disjointness (Section \ref{sec:disj}). In Section
\ref{sec:pointer-jumping} we give our new upper bound (Section
\ref{sec:pj-upper}) and quantum lower bound (Section
\ref{sec:pj-lower}) for the pointer-jumping problem.

\subsection*{Subsequent Results}

Subsequent to the publication of the preliminary version of this
paper~\cite{KNTZ01} several new related results have appeared.  First,
Razborov proves in Ref.~\cite{R03} that the quantum
communication complexity of the Set Disjointness problem is indeed
$\Omega(\sqrt{N})$, no matter how many rounds are allowed. An upper
bound of $O(\sqrt N)$ is given by Aaronson and Ambainis~\cite{AA05}. A
result by Jain, Radhakrishnan, and Sen in Ref.~\cite{JRS03} shows that
the complexity of protocols solving this problem in $k$ rounds is at
least $\Omega(n/k^2)$. The same authors show in Ref.~\cite{JRS02a}
that quantum protocols with $k-1$ rounds for the Pointer Jumping
function $f_k$ have complexity $\Omega(n/k^4)$, but this result seems
to hold only for the case of protocols {\em without\/} prior
entanglement. The same authors~\cite{JRS02b} also consider the
complexity of quantum protocols for the version of the Pointer Jumping
function, in which not only one bit of the last vertex has to be
computed, but its full name.  Several papers
(\cite{SV01,K04,JRS03,JRS02a,CR04}) have used the information
theoretic techniques developed in the present paper.

In this paper, we improve the dependence of communication complexity
lower bounds on the number of rounds, as compared to our results in
Ref.~\cite{KNTZ01}.  To achieve this, we use a different
information-theoretic tool based on the quantum Hellinger
distance. The version of our Average Encoding Theorem based on
Hellinger distance was independently found by Jain
\etal{}~\cite{JRS03}.

\section{Information Theory Background}
\label{sec:info-background}

The quantum mechanical analogue of a random variable is a
probability distribution over superpositions, also called a {\em
mixed state}. For the mixed state~$X = \{p_i,\ket{\phi_i}\}$,
where $\ket{\phi_i}$ has probability~$p_i$, the {\em density
matrix\/} is defined as~$\brho_X  = \sum_i p_i
\ketbra{\phi_i}{\phi_i}$. Density matrices are Hermitian, positive
semi-definite, and have trace $1$. I.e., a density matrix has an
eigenvector basis, all the eigenvalues are real and between zero
and one, and they sum up to one.

\subsection{Trace Norm And Fidelity}

The {\em trace norm\/} of a matrix~$A$ is defined as~$\trn{A} =
\trace\,{\sqrt{A^\dagger A}}$, which is the sum of the magnitudes
of the singular values of~$A$. Note that if $\rho$ is a density
matrix, then it has trace norm one. If $\phi_1,\phi_2$ are pure
states then:
 $$
\trnorm{\density{\phi_1} - \density{\phi_2}}
    ~~=~~ 2 \sqrt{ 1 - \size{\braket{\phi_1}{\phi_2}}^2}.
$$

We will need the following consequence of Kraus representation
theorem (see for example Preskill's lecture notes~\cite{P98}):

\begin{lemma}
\label{lem:superop-dec} For each Hermitian matrix $\rho$ and each
trace-preserving completely positive superoperator $T$:
$\|T(\rho)\|_{\rm t} \le \|\rho\|_{\rm t}$.
\end{lemma}

A useful alternative to the trace metric as a measure of closeness
of density matrices is {\em fidelity}. Let $\brho$ be a mixed
state with support in a Hilbert space~$\aitch$. A {\em
purification\/} of $\brho$ is any pure state~$\ket{\phi}$ in an
extended Hilbert space~$\aitch \tensor \kay$ such
that~$\trace_{\kay} \ketbra{\phi}{\phi} = \brho$. Given two
density matrices~$\brho_1,\brho_2$ on the same Hilbert
space~$\aitch$, their {\em fidelity\/} is defined as
$$
F(\brho_1,\brho_2) \;\;=\;\; \sup\,
\size{\braket{\phi_1}{\phi_2}}^2,
$$

where the supremum is taken over all purifications~$\ket{\phi_i}$
of~$\brho_i$ in the same Hilbert space. Jozsa~\cite{Jozsa94} gave
a simple proof, for the finite dimensional case, of the following
remarkable equivalence first established by
Uhlmann~\cite{Uhlmann76}.

\begin{fact}[Jozsa]
\label{fact:jozsa-short} For any two density matrices $\brho_1,
\brho_2$ on the same finite dimensional space~$\aitch$,
$$
F(\brho_1,\brho_2) \;\;=\;\; \left[\trace\left(
                             \sqrt{{\brho_1}^{1/2}\,\brho_2\,{\brho_1}^{1/2}}
                             \right) \right]^2
                   \;\;=\;\; \trn{ \sqrt{\brho_1}\sqrt{\brho_2}}^2.
$$
\end{fact}

Using this equivalence, Fuchs and van de Graaf~\cite{FG99} relate
fidelity to the trace distance.

\begin{fact}[Fuchs, van de Graaf]
\label{fact:fidelity} For any two mixed states $\brho_1,\brho_2$,
$$
1 - \sqrt{F(\brho_1,\brho_2)} \;\; \le \;\; \frac{1}{2}
\trn{\brho_1 - \brho_2} \;\;\le\;\;
 \sqrt{1 - F(\brho_1,\brho_2)}.
$$
\end{fact}

While the definition of fidelity uses purifications of the mixed
states and relates them via the inner product, fidelity can also be
characterized via measurements (see Nielsen and Chuang~\cite{NC00}).\\

\begin{fact}\label{fact:fidelitymeasurement} 
For two probability distributions~$p,q$ on finite sample spaces,
let~$F(p,q) = (\sum_i \sqrt{ p_i q_i} )^2$ denote their
fidelity. Then, for any two mixed states~$\rho_1, \rho_2$,
\[
F(\brho_1,\brho_2) ~~=~~ \min_{\{E_m\}} F(p_m,q_m),
\]
where the minimum is over all POVMs~$\{E_m\}$, and $p_m=\trace(\brho_1
E_m), q_m=\trace(\brho_2 E_m)$ are the probability distributions
created by the measurement on the states.
\end{fact}

A useful property of the trace distance $\trn{\brho_1-\brho_2}$ as a
measure of distinguishability is that it is a metric, and hence
satisfies the triangle inequality. This is not true for fidelity
$F(\brho_1,\brho_2)$ or for $1-F(\brho,\brho_2)$.  Fortunately, a
variant of fidelity is actually a metric. Denote by
$$
\hel{\brho_1,\brho_2} ~~=~~ \sqrt{ 1-\sqrt{ F(\brho_1,\brho_2) } }
$$
the quantum {\em Hellinger distance\/}. Clearly
$\hel{\brho_1,\brho_2}$ inherits most of the desirable properties
of fidelity, like unitary invariance, definability as a maximum
over all measurements of the classical Hellinger distance of the
resulting distributions, and so on. To see that
$\hel{\brho_1,\brho_2}$ is actually a metric one can simply use
Fact~\ref{fact:fidelitymeasurement} to reduce this problem to
showing that the classical Hellinger distance is a metric, which
is well known.

Analogously to Lemma~\ref{lem:superop-dec}, due to the
monotonicity of fidelity~\cite{NC00}, we have:

\begin{lemma}
\label{lem:superop-dec-hell} For all density matrices
$\rho_1,\rho_2$ and each trace-preserving completely positive
superoperator $T$: $\hel{T(\rho_1),T(\rho_2)} \le
\hel{\rho_1,\rho_2}$.
\end{lemma}

 Let us also note the following relation between the Hellinger
distance and the trace norm that follows directly from
Fact~\ref{fact:fidelity}.

\begin{lemma}\label{lem:tracevshellinger}
For any two mixed states $\brho_1,\brho_2$,
$$\hels{\brho_1,\brho_2} \;\; \le \;\; \frac{1}{2} \trn{\brho_1 -
\brho_2} \;\;\le\;\; \sqrt{2}\cdot\hel{\brho_1,\brho_2}.
$$
\end{lemma}

We will sometimes work with $\hels{\cdot,\cdot}$ instead of $\hel{\cdot,\cdot}$. This is not a metric, but
it is true that for all density matrices $\rho_1,\rho_2,\rho_3$:
\[\hels{\rho_1,\rho_2}\le\left(\hel{\rho_1,\rho_3}+\hel{\rho_3,\rho_2}\right)^2
\le 2\hels{\rho_1,\rho_3}+2\hels{\rho_3,\rho_2}.\]

\subsection{Local Transition Between Bipartite States}
\label{sec:local}

Jozsa~\cite{Jozsa94} proved:

\begin{theorem}[Jozsa]
\label{thm:jozsa} Suppose~$\ket{\phi_1}, \ket{\phi_2} \in \aitch
\tensor \kay$ are the purifications of two density
matrices~$\brho_1,\brho_2$ in~$\aitch$. Then, there is a local
unitary transformation~$U$ on~$\kay$ such that $F(\brho_1,
\brho_2) = \size{\bra{\phi_1} (I\tensor U)\ket{\phi_2}}^2$.
\end{theorem}

As noticed by Lo and Chau~\cite{LC98} and Mayers~\cite{M97},
Theorem \ref{thm:jozsa} immediately implies that if two states
have close reduced density matrices, than there exists a {\em
local} unitary transformation transforming one state close to the
other. Formally,

\begin{lemma}
(Local Transition Lemma, based on Refs.~\cite{LC98,M97,Jozsa94,FG99}) \label{lem:close}
Let~$\brho_1,\brho_2$ be two mixed states with support in a
Hilbert space $\aitch$. Let $\kay$ be any Hilbert space of
dimension at least~$\dim(\aitch)$, and $\ket{\phi_i}$ any
purifications of~$\brho_i$ in $\aitch \tensor \kay$.

Then, there is a local unitary transformation~$U$ on~$\kay$ that
maps~$\ket{\phi_2}$ to~$\ket{\phi'_2} = I\tensor U\ket{\phi_2}$
such that
$$\hel{ \ketbra{\phi_1}{\phi_1} , \ketbra{\phi'_2}{\phi'_2}}
\;\;=\;\; \hel{\brho_1,\brho_2}.$$

Furthermore,
$$\trn{ \ketbra{\phi_1}{\phi_1} - \ketbra{\phi'_2}{\phi'_2}}
\;\;\le\;\; 2 \trn{\brho_1-\brho_2}^{\frac{1}{2}}.$$
\end{lemma}

\begin{proof}(Of Lemma~\ref{lem:close}):
By Theorem~\ref{thm:jozsa}, there is a (local) unitary
transformation~$U$ on~$\kay$ such that~$(I\tensor U)\ket{\phi_2} =
\ket{\phi'_2}$, a state which achieves fidelity: $F(\brho_1,
\brho_2) = \size{\braket{\phi_1}{\phi'_2}}^2$. Hence the statement
about the Hellinger distance holds.


By Lemma~\ref{lem:tracevshellinger}
\begin{eqnarray*}
\lefteqn{\trn{ \ketbra{\phi_1}{\phi_1} - \ketbra{\phi'_2}{\phi'_2}}} \\
&\le& 2\sqrt2\cdot \hel{\ketbra{\phi_1}{\phi_1}, \ketbra{\phi'_2}{\phi'_2}}\\
&=& 2\sqrt2\cdot\hel{\brho_1,\brho_2}\\
&\le& 2\cdot\trn{\brho_1-\brho_2}^{\frac{1}{2}}.
\end{eqnarray*}

\end{proof}

\subsection{Entropy, Mutual Information, And Relative Entropy.}
\label{sec:info-background:entropy}

$H(\cdot)$ denotes the binary entropy function $H(p)=p \log({1 \over
p})+ (1-p) \log({ 1 \over 1-p})$. The {\em Shannon entropy\/}~$S(X)$
of a classical random variable~$X$ on a finite sample space is~$\sum_x
p_x \log({1 \over p_x})$ where $p_x$ is the probability the random
variable $X$ takes value $x$. The {\em mutual information\/}~$I(X:Y)$
of a pair of random variables~$X,Y$ is defined to be
$I(X:Y)=H(X)+H(Y)-H(X,Y)$. For other equivalent definitions, and more
background on the subject see, e.g., the book by Cover and
Thomas~\cite{CT91}.

We use a simple form of Fano's inequality.
\begin{fact}[Fano's inequality]
\label{cl:mut} Let~$X$ be a uniformly distributed Boolean random
variable, and let~$Y$ be a Boolean random variable such
that~$\prob(X=Y) = p$. Then~$I(X:Y) \ge 1 - H(p)$.
\end{fact}


The Shannon entropy and the mutual information functions have
natural generalizations to the quantum setting. The {\em von
Neumann entropy\/}~$S(\brho)$ of a density matrix~$\brho$ is
defined as~$S(\brho) = - \trace\,\brho\log\brho = - \sum_i
\lambda_i \log \lambda_i$, where~$\{\lambda_i\}$ is the multi-set
of all the eigenvalues of~$\brho$. Notice that the eigenvalues of
a density matrix form a probability distribution. In fact, we can
think of the density matrix as a mixed state that takes the $i$'th
eigenvector with probability $\lambda_i$. The von Neumann entropy
of a density matrix $\brho$ is, thus, the entropy of the classical
distribution $\brho$ defines over its eigenstates.

The mutual information~$I(X:Y)$ of two disjoint quantum systems
$X,Y$ is defined to be $I(X:Y)=S(X)+S(Y)-S(XY)$, where~$XY$ is the
density matrix of the system that includes the qubits of both
systems. Then

\begin{eqnarray}
\label{eqn:infeq}
I(X:YZ) & = & I(X:Y)+I(XY:Z)-I(Y:Z), \\
\label{eqn:infineq}
I(X:YZ) & \ge  & I(X:Y),
\end{eqnarray}

Equation~(\ref{eqn:infineq}) is in fact equivalent to the {\em
strong sub-additivity property\/} of von Neumann entropy.

We need the following slight generalization of Theorem 2 in Cleve
\etal{}~\cite{CDNT98}.

\begin{lemma}\label{lem:factor2}
Let Alice own a state $\rho_A$ of a register $A$. Assume Alice and
Bob communicate and apply local transformations, and at the end
register $A$ is measured in the standard basis. Assume Alice sends
Bob at most $k$ qubits, and Bob sends Alice arbitrarily many
qubits. Further assume all these local transformations do not
change the state of register $A$, if $A$ is in a classical state.
Let $\rho_{AB}$ be the final state of $A$ and Bob's private qubits
$B$. Then $I(A:B)\le 2k$.
\end{lemma}

\begin{proof}
Considering the joint state of register $A$ and Bob's qubits,
there cannot be any interference between basis states differing on
$A$. Thus we can assume that $\rho_A$ is measured in the
beginning, i.e., that $\rho_A$ is classical. In this case the
result directly follows from Theorem 2 in Ref.~\cite{CDNT98}.
\end{proof}

Note that in the above lemma Alice and Bob can use Bob's free
communication to set up an arbitrarily large amount of
entanglement independent of $\rho_A$.

The {\em relative\/} von Neumann entropy of two density matrices,
defined by $S(\rho\|\sigma)=
\trace\,\rho\log\rho-\trace\,\rho\log\sigma$. One useful fact to know
about the relative entropy function is that $I(A:B)=
S(\rho_{AB}\|\rho_A\otimes \rho_B)$. For more properties of this
function see Refs.~\cite{P98,NC00}.

\section{Informational Distance And New Lower Bounds On Relative
Entropy} \label{sec:information-theory-new}

\subsection{New Lower Bounds On Relative Entropy}
\label{sec:relative-trace}

We now prove that the relative entropy $S(\rho_1\|\rho_2)$ is
lower bounded by $\Omega(\trn{\rho_1-\rho_2}^2)$ and by
$\Omega(h^2(\rho_1,\rho_2))$. We believe these results are of
independent interest. A classical version of the theorem can be
found in, e.g., Cover and Thomas' book on Information Theory~\cite{CT91}.

\begin{theorem}
\label{thm:relative-trace} For all density matrices
$\rho_1,\rho_2$:
\[S(\rho_1\|\rho_2) ~~\ge~~ \frac{1}{2\ln2}\trn{\rho_1-\rho_2}^2.\]
\end{theorem}

Although this relationship has appeared in the literature~\cite{OP93},
it was rediscovered by several authors, including us.  Below we give a
proof of this theorem for completeness. The earlier version of our
paper~\cite{KNTZ01} contained a more complicated proof.

\begin{proof}(Theorem~\ref{thm:relative-trace})
The proof goes by reduction to the classical case. Consider the
classical distributions~$\tilde{\rho}_1, \tilde{\rho}_2$ obtained by
measuring~$\rho_1, \rho_2$ in the basis diagonalizing their
difference~$\rho_1 - \rho_2$. It is known~\cite{P98,NC00} that
\begin{eqnarray*}
\norm{\tilde{\rho}_1 - \tilde{\rho}_2}_1 & = & \trn{\rho_1-\rho_2}.
\end{eqnarray*}
Due to Lindblad-Uhlmann monotonicity of
relative von Neumann entropy~\cite{P98,NC00}, 
\begin{eqnarray*}
S(\rho_1\|\rho_2) & \geq & S(\tilde{\rho}_1 \| \tilde{\rho}_2).
\end{eqnarray*}
The classical version of the theorem~\cite{CT91} now gives
\begin{eqnarray*}
S(\tilde{\rho}_1 \| \tilde{\rho}_2) 
    & \geq & \frac{1}{2 \ln 2} \norm{\tilde{\rho}_1 - \tilde{\rho}_2}_1^2 \\
    & = &  \frac{1}{2 \ln 2} \trn{\rho_1-\rho_2}^2.
\end{eqnarray*}
This completes the proof.
\end{proof}

Now we show an analogous result for the quantum Hellinger distance.

\begin{theorem}
\label{thm:relative-hell} For all density matrices
$\rho_1,\rho_2$:
\[S(\rho_1\|\rho_2) ~~\ge~~ \frac{2}{\ln 2}\,\hels{\rho_1,\rho_2}.\]
\end{theorem}

This theorem has also been shown independently by Jain \etal{}~\cite{JRS03}.

\begin{proof}
We first show that the theorem holds when $\rho_1$ and $\rho_2$
are classical distributions, and then generalize this to the
quantum case.

In the classical case we first show $S(\rho_1\|\rho_2)\ge
-2\log(1-\hels{\rho_1,\rho_2})$. This was shown by Dacunha-Castelle in
Ref.~\cite{D78}.

\begin{eqnarray*}
\log(1-\hels{\rho_1,\rho_2})
&=&\log(\sqrt{F(\rho_1,\rho_2)})\\
&=&\log\left(\sum_i \sqrt{\rho_1(i)\rho_2(i)}\right)\\
&=&\log\left(\sum_i \rho_1(i)\frac{\sqrt{\rho_2(i)}}{\sqrt{\rho_1(i)}}\right)\\
&\ge&\sum_i\rho_1(i)\log\left(\frac{\sqrt{\rho_2(i)}}{\sqrt{\rho_1(i)}}\right)\\
&=&-\frac{1}{2} S(\rho_1\|\rho_2) .
\end{eqnarray*}
The first equation is by definition of $h$, the second by definition
of the classical fidelity function, and the inequality is by an
application of Jensen's inequality.

Having that, $S(\rho_1\|\rho_2)\ge \frac{2}{\ln 2}
\hels{\rho_1,\rho_2}$ using $-\ln (1-x)\ge x$ for all $0\le x\le
1$ and so the theorem holds in the classical case.

To show the quantum case recall that both $\hel{\cdot,\cdot}$ and
$S(\cdot\|\cdot)$ can be defined as the maximum over all POVM
measurements of the classical versions of these functions on the
distributions obtained by the measurements. Fix a POVM $\{E_m\}$ that
maximizes $\hel{p,q}$ for the distributions $p,q$ obtained
from $\rho_1,\rho_2$. Then $S(\rho_1\|\rho_2)\ge S(p\|q)$ by
Lindblad-Uhlmann monotonicity, and~$S(p\|q) \ge \frac{2}{\ln 2}
\hels{p,q} = \frac{2}{\ln 2} \hels{\rho_1,\rho_2}$ because
$\hel{p,q}= \hel{\rho_1,\rho_2}$. The result follows.
\end{proof}

\subsection{Informational Distance}
\label{sec:informational-distance}

{}From Theorem~\ref{thm:relative-hell} follows that for a
bipartite state $\rho_{AB}$,
\[
I(A:B) ~~=~~ S(\rho_{AB}\|\rho_A\otimes \rho_B)
       ~~\ge~~ \frac{2}{\ln2}\hels{\rho_{AB},\rho_A\otimes\rho_B}.
\]
Thus the distance between the tensor product state and the
``real'' (possibly entangled) bipartite state can be bounded in
terms of the Hellinger distance. We call the quantity
$D(A:B)=\hel{\rho_{AB},\rho_A\otimes\rho_B}$ the ``informational
distance.'' $D(A:B)$ measures the amount of correlation between
the quantum registers $A$ and $B$, and can be positive even when
the system is classical or not entangled. Later we state some of
its properties and use it for proving the quantum communication
lower bound on the pointer jumping problem.

The next lemma collects a few immediate properties of
informational distance.

\begin{lemma}
\label{lem:inf-dist} For all states $\rho_{XYZ}$ the following
hold:
\begin{enumerate}
\item
\label{enu-sym} $D(X:Y)=D(Y:X)$,
\item
\label{enu-02} $0\le D(X:Y)\le 1$,
\item
\label{enu-F} $D(X:Y)\ge \hel{T(\rho_{XY}),T (\rho_X\otimes
\rho_Y)} $ for all completely positive, trace-preserving
superoperators $T$,
\item
\label{enu-measurement} $D(XY:Z)\ge D(X:Z)$,
\item
\label{enu-thm} $D(X:Y)\le \sqrt{I(X:Y)}$.
\end {enumerate}
\end{lemma}

\begin{proof}
(\ref{enu-sym}) is true by definition, (\ref{enu-02}) follows from
the definition and the triangle inequality,
(\ref{enu-F},\ref{enu-measurement}) follow from
Lemma~\ref{lem:superop-dec-hell} and (\ref{enu-thm}) from
Theorem~\ref{thm:relative-hell}.
\end{proof}

We now examine the informational distance in the special case
where $\rho_{QX}$ is block diagonal, with classical $\rho_X$. We
denote by~$\rho_Q^{(x)}$ the density matrix obtained by fixing $X$ to
some classical value $x$ and normalizing.  $\Pr(x)$ is the
probability of $X=x$.

\begin{lemma}
\label{lem:propID} For all block diagonal $\rho_{QX}$, where
$\rho_X$ corresponds to a classical distribution,
\begin{enumerate}
\item
\label{enu:GeneralX}
  $D^2(Q:X)=\E_x \; \hels{ \rho_Q^{(x)},\rho_Q}.$
\item
\label{enu:BooleanX}
  Further assume $X$ is Boolean with $\Pr(X=1)=\Pr(X=0)=1/2$. Let
  there be a measurement acting on the $Q$ system only, yielding a
  Boolean random variable $Y$ with $\Pr(X=Y)\ge 1-\epsilon$ and
  $\Pr(X\neq Y)\le\epsilon$.
  Then $D^2(Q:X)\ge 1/8-\epsilon/2$.
\end{enumerate}
\end{lemma}
The first item is true because $\rho_{QX}$ is block-diagonal with
respect to $X$.  In the second item, notice that the same measurement
applied to $\rho_X\otimes \rho_Q$ yields a distribution with
$\Pr(X=Y)=\Pr(X\neq Y)=1/2$, because $Q$ is independent of $X$, and $X$
is uniform.  Observe that~$\trn{\rho_{XQ}-\rho_X \otimes
\rho_Q}\ge\trn{\rho_{XY}-\rho_X\otimes\rho_Y}\ge 1-2\epsilon$ and then
apply Lemma~\ref{lem:tracevshellinger}. Note that this is a rather
crude estimate, since $D(Q:X)$ approaches $1-1/\sqrt 2$ when
$\epsilon$ goes to zero.

\subsection{The Average Encoding Theorem}
\label{sec:average-encoding}

A corollary of Theorems~\ref{thm:relative-trace},\ref{thm:relative-hell} is the following
``Average encoding theorem'':

\begin{theorem}[Average encoding theorem]
\label{thm:average-encoding} Let $x \mapsto \rho_x$ be a quantum
encoding mapping~an $m$ bit string~$x\in \set{0,1}^m$ into a mixed
state with density matrix $\rho_x$. Let~$X$ be distributed
over~$\set{0,1}^m$, where $x\in\set{0,1}^m$ has probability $p_x$,
let~$Q$ be the encoding of~$X$ according to this map, and
let~$\bar{\rho} = \sum_x p_x\rho_x$. Then,
\begin{eqnarray*}
\sum_x p_x\trn{\bar{\rho} - \rho_x}
    & \le & \left[(2 \ln 2) \; I(Q:X) \right]^{1/2}
\end{eqnarray*}
and
\begin{eqnarray*}
\sum_x p_x \;\; \hels{\bar{\rho} , \rho_x}
    & \le & \frac{\ln 2}{2} \; I(Q:X) .
\end{eqnarray*}

\end{theorem}

In other words, if an encoding~$Q$ is only weakly correlated to a
random variable~$X$, then the ``average encoding''~$\bar{\rho}$ is
in expectation (over a random string) a good approximation of any
encoded state. Thus, in certain situations, we may dispense with
the encoding altogether, and use the single state~$\bar{\rho}$
instead. The preliminary version of our paper~\cite{KNTZ01} did
not include the second statement. The present stronger version was
also observed independently by Jain \etal{}~\cite{JRS03}.

\begin{proof}(Of Theorem~\ref{thm:average-encoding})
In the setting of the Average encoding theorem we have a random
variable that is distributed over~$\set{0,1}^m$, and a quantum
encoding $x \mapsto \rho_x$ mapping~$m$ bit strings~$x\in
\set{0,1}^m$ into mixed states with density matrices $\rho_x$.
Let~$X$ be the register holding the input $x$ and $Q$ be the
register holding the encoding. Let us also define the average
encoding $\bar{\rho} = \sum_x p_x\rho_x$.

Then, by Theorem \ref{thm:relative-trace},

\begin{eqnarray*}
I(Q:X) &=& S(\rho_{QX}\|\rho_{Q}\otimes \rho_X) ~~\ge~~ {1 \over 2 \ln
2} \trn{\rho_{QX}-\rho_{Q} \otimes \rho_X}^2
\end{eqnarray*}

The density matrix $\rho_X$ of the $X$ register alone is diagonal
and contains the values $p_x$ on the diagonal, the density matrix
$\rho_Q$ of the $Q$ register alone is $\bar{\rho}$, and the
density matrix $\rho_Q \otimes \rho_X$ is block diagonal and the
$x$'th block is of the form $p_x \bar{\rho}$. Also, the density
matrix $\rho_{QX}$ of the whole system is block diagonal, with
$p_x \rho_x$ in the $x$'th block. Thus, $\trn{\rho_{QX}-\rho_{Q}
\otimes \rho_X} =\sum_x p_x \trn{\rho_x-\bar{\rho}}$, and so $\E_x
\trn{\rho_x-\bar{\rho}} \le \sqrt{2 \ln 2} \sqrt{I(Q:X)}$.

The second statement follows analogously using Theorem~\ref{thm:relative-hell}.
\end{proof}


\section{The Communication Complexity Model}
\label{sec:cc-model}

In the quantum communication complexity model~\cite{Y93}, two
parties Alice and Bob hold qubits. When the game starts Alice
holds a classical input~$x$ and Bob holds~$y$, and so the initial
joint state is simply~$\ket{x} \tensor \ket{y}$. Furthermore each
player has an arbitrarily large supply of private qubits in some
fixed basis state. The two parties then play in turns. Suppose it
is Alice's turn to play. Alice can do an arbitrary unitary
transformation on her qubits and then send one or more qubits to
Bob. Sending qubits does not change the overall superposition, but
rather changes the ownership of the qubits, allowing Bob to apply
his next unitary transformation on the newly received qubits.
Alice may also (partially) measure her qubits during her turn. At
the end of the protocol, one player makes a measurement and
declares the result of the protocol. In a classical probabilistic
protocol the players may only exchange classical messages.

In both the classical and quantum settings we can also define a
public coin model. In the classical public coin model the players
are also allowed to access a shared source of random bits without
any communication cost. The classical public and private coin
models are strongly related~\cite{KN97}. Similarly, in the quantum
public coin model Alice and Bob initially share an arbitrary
number of quantum bits which are in some pure state that is
independent of the inputs. This is better known as communication
with prior entanglement~\cite{CDNT98,BW01}.

The complexity of a quantum (or classical) protocol is the number of qubits
(respectively, bits) exchanged between the two players.
We say a protocol {\em computes\/}
a function~$f : \X \times \Y \mapsto \{0,1\}$
with~$\epsilon \ge 0$ error if, for any input~$x \in \X,y \in \Y$,
the probability
that the two players compute~$f(x,y)$ is at least $1-\epsilon$.
$Q_\epsilon(f)$
(resp.~$R_\epsilon(f)$) denotes the complexity of the best quantum
(resp.~probabilistic)
protocol that computes~$f$ with at most~$\epsilon$ error.
For a player~$P \in \set{{\rm Alice},\;{\rm Bob}}$,
$Q^{c,P}_{\epsilon}(f)$ denotes
the complexity of the best quantum protocol that
computes~$f$ with at most~$\epsilon$ error
with only~$c$ messages (called rounds in the literature),
where the first message is sent by~$P$.
If the name of the player is omitted from the superscript,
either player is allowed to start the protocol.
We say a protocol~$\cp$ {\em computes\/}~$f$ with~$\epsilon$ error
with respect to a distribution~$\mu$ on~$\X \times \Y$,
if
$$\prob_{(x,y) \in \mu, \cp}( \cp(x,y)=f(x,y)) \;\;\ge\;\; 1-\epsilon.$$
$Q^{c,P}_{\mu,\epsilon}(f)$ is the complexity of computing~$f$
with at most~$\epsilon$ error with respect to~$\mu$, with only~$c$
messages where the first message is sent by player~$P$. We will
use the notation~$\eqcc$ (rather than~$Q^*$, as in the literature)
for communication complexity in the public coin model. In all the
above definitions, we may replace $\mu$ with $U$ when $\mu$ is the
uniform distribution over the inputs.

The following is immediate.

\begin{fact}
For any distribution~$\mu$, number of messages~$c$ and
player~$P$,~$\eqcc^{c,P}_{\mu,\epsilon}(f) \le
Q^{c,P}_{\mu,\epsilon}(f) \le Q^{c,P}_{\epsilon}(f)$.
\end{fact}

We put two constraints on protocols in the above definitions:
\begin{itemize}
\item
We assume that the two players do not modify the qubits holding
the classical input during the protocol. This does not affect the
aspect of communication we focus on in this paper.
\item
We demand that the length of the $i$'th message sent in a protocol is
known in advance.  This restriction is also implicit in Yao's
definition of quantum communication complexity using interacting
quantum circuits~\cite{Y93}.
\end{itemize}

To illustrate this, think of a public coin classical protocol in
which Alice first looks at a public coin and if the coin is ``head''
sends in the first round a message of $c$ qubits and in the second
round a message of 1 qubit, otherwise she sends one qubit in the
first round and $c$ qubits in the second. In such a protocol the
number of message bits sent in the first round is not known in
advance, and so such a protocol is not allowed in our model.

A $k$ round protocol with communication complexity $c$ in the more
general model, in which the restriction above is absent, can be
simulated in our model losing a factor of $k$ in the communication
complexity. To show this one invokes the principle of safe
storage. The principle says that instead of a mixed state depending on
measurement results, we may have a superposition over the measurement
results and the messages. Note that in such a superposition there may
be messages of different lengths (augmented by some blanks). In the
worst case, the length of a single message is now $c$, so the overall
communication cost is at most $kc$, and the number of rounds used is
always the worst case number of rounds. In the example above we get a
$2c$ communication complexity.

\section{The Role Of Interaction In Quantum Communication}
\label{sec:Sk}

In this section, we prove that allowing more interaction between
two players in a quantum communication game can substantially
reduce the amount of communication required. In Section
\ref{sec:problem} we define a communication problem and formally
state our results (giving an overview of the proof), then in
Section \ref{sec:Sk-key-lemmas} we give the details of the proofs.
For the most part, we will concentrate on communication in a
constant number of rounds. Section~\ref{sec:disj} describes the
application to the disjointness problem.
Section~\ref{sec:non-const} discusses our results in the case
where the number of messages grows as a function of the input
size.

\subsection{The Communication Problem And Its Complexity}
\label{sec:problem}

We define a sequence of problems~$S_1,S_2,\ldots,S_k,\ldots$ by
induction. The problem~$S_1$ is the index function, i.e., Alice
has an~$n$-bit string~$x \in \X_1 = \set{0,1}^n$, Bob has an
index~$i \in \Y_1 = [n]$ and the desired output is~$S_1(x,i)=x_i$.
Suppose we have already defined the function~$S_{k-1}: \X_{k-1}
\times  \Y_{k-1} \to \set{0,1}$. In the problem~$S_{k}$, Alice has
as input her part of~$n$ independent instances of~$S_{k-1}$, i.e.,
$x \in \X_{k-1}^n$, Bob has his share of~$n$ independent instances
of~$S_{k-1}$, i.e., $y \in \Y_{k-1}^n$, and in addition, there is
an extra input~$a \in [n]$ which is given to Alice if~$k$ is even
and to Bob if~$k$ is odd. The output we seek is the solution to
the~$a$'th instance of~$S_{k-1}$. In other words,
$S_{k}(x_1,\ldots,x_n,a,y_1,\ldots,y_n)=S_{k-1}(x_a,y_a)$.

Note that the size of the input to the problem~$S_k$ is~$N =
\Theta(n^k)$. If we allow~$k$ message exchanges for solving the
problem, it can be solved by exchanging~$\Theta(\log N) =
\Theta(k\log n)$ bits: for~$k = 1$, Bob sends Alice the index~$i$
and Alice then knows the answer; for~$k > 1$, the player with the
index~$a$ sends it to the other player and then they recursively
solve for~$S_{k-1}(x_a,y_a)$. However, we show that if we allow
one less message, then no quantum protocol can compute~$S_k$ as
efficiently. In fact, no quantum protocol can compute the function
as efficiently even if we allow error, and only require small
probability of error on average.

\begin{theorem}
\label{thm:lb} For all constant~$k \ge 1$ and $0 \le \epsilon <
\frac{1}{2}$ we have
$$ \eqcc^k_{U,\epsilon}(S_{k+1})
    \;\;=\;\; \Omega\left( N^{1/(k+1)} \right).
$$
\end{theorem}

To prove this theorem we prove a stronger intermediate claim.
Let~$P_1$ be Bob, and for~$k \ge 2$, let~$P_k$ denote the player
that holds the index~$a$ in an instance of~$S_k$ ($a$ indicates
which of the~$n$ instances of~$S_{k-1}$ to solve). Let~$\bar{P}_k$
denote the other player. We refer to~$\bar{P}_k$ as the ``wrong''
player to start a protocol for~$S_k$. The stronger claim is that
any~$k$ message protocol for~$S_k$ in which the wrong player
starts is exponentially inefficient as compared to the~$\log N$
protocol described above.

\begin{lemma}
\label{lem:lbi} For all constant~$k \ge 1$ and $0 \le \epsilon <
\frac{1}{2}$ we have $ \eqcc^{k,\bar{P}_k}_{U,\epsilon}(S_k)
    \;\;=\;\; \Omega(n) \;\;=\;\; \Omega\left( N^{1/k} \right).
$
\end{lemma}

Indeed, there is a classical~$k$-message, $O(n)$-bit
protocol in which the wrong
player starts, so our lower bound is optimal.

Theorem~\ref{thm:lb} now follows directly.

\begin{proof}(Of Theorem~\ref{thm:lb}):
It is enough to show the lower bound for the two cases when the protocol
starts either with~$P_{k+1}$ or with the other player.

Let~$P_{k+1}$ be the player to start. Note that if we set~$a$ to a
fixed value, say~$1$, then we get an instance of~$S_k$.
So~$\eqcc^{k,P_{k+1}}_{U,\epsilon}(S_k) \le
\eqcc^{k,P_{k+1}}_{U,\epsilon}(S_{k+1})$. But~$P_{k+1} =
\bar{P}_k$, so the bound of Lemma~\ref{lem:lbi} applies.

Let player~$\bar{P}_{k+1}$ be the one to start. Then, observe that
if we allow one more message (i.e.,~$k+1$ messages in all), the
complexity of the problem only decreases:
$\eqcc^{k+1,\bar{P}_{k+1}}_{U,\epsilon}(S_{k+1}) \le
\eqcc^{k,\bar{P}_{k+1}}_{U,\epsilon}(S_{k+1})$. So we again get
the bound from Lemma~\ref{lem:lbi}.
\end{proof}

We prove Lemma~\ref{lem:lbi} by induction. First, we show that the
index function is hard to solve with one message if the wrong
player starts. This essentially follows from the lower bound for
random access codes~\cite{Nayak99,AmbainisNTV02}. The only difference
is that we seek a lower bound for a protocol that has low error
probability {\em on average\/} rather than in the worst case, so
we need a refinement of the original argument. We give this in the
next section.

\begin{lemma}
\label{lem:k=1} For any~$0 \le \epsilon \le 1$ we have
$\eqcc^{1,A}_{U,\epsilon}(S_1) \ge {1 \over 2} (1-H(\epsilon)) n$.
\end{lemma}

Next, we show that if we can solve~$S_k$ with~$k$ messages with the
wrong player starting, then we can also solve~$S_{k-1}$ with
only~$k-1$ messages of smaller total length, again with the wrong
player starting, at the cost of a slight increase in the
average probability of error.

\begin{lemma}
\label{lem:reduction} 
For ~$k \ge 2$ and $0 \le \epsilon < \frac{1}{2}$, let~$\cp$ be any
protocol that solves~$S_{k}$ with respect to the uniform
distribution~$U$ with error~$\epsilon$, and~$k$ messages starting
with~$\bar{P}_k$. Let the communication complexity of~$\cp$ be~$\ell =
\ell_1 + \bar{\ell}$ with $\ell_1$ being the length of the first
message sent.
Then, $
\eqcc^{k-1,\bar{P}_{k-1}}_{U,\epsilon'}(S_{k-1}) \;\le\; \bar{\ell}, $
where~$\epsilon'=\epsilon + 2 (\ell_1/n)^{1/2}$.
\end{lemma}
We defer the proof of this lemma to a later section, but show how
it implies Lemma~\ref{lem:lbi} above.

\begin{proof}(Of Lemma~\ref{lem:lbi}):
We prove the lemma by induction on~$k$. The case~$k=1$ is handled by
Lemma~\ref{lem:k=1}. Suppose the statement holds for~$k - 1$. We prove
by contradiction that it holds for~$k$ as well. If~$\ell =
\eqcc^{k,\bar{P}_k}_{U,\epsilon}(S_{k}) = o(n)$, then by
Lemma~\ref{lem:reduction} there is a~$k-1$ message protocol
for~$S_{k-1}$ with the wrong player starting, with error~$\epsilon' =
\epsilon + o(1) < \frac{1}{2}$, and with communication
complexity at most~$\ell = o(n)$. This contradicts the induction
hypothesis.
\end{proof}

\subsection{The Key Lemmas}
\label{sec:Sk-key-lemmas}

We now prove average case hardness of the index function.

\begin{proof}(Of Lemma~\ref{lem:k=1}):
Consider any protocol for~$S_1$ with Alice sending the first (and
only) message. Let~$\epsilon_i$ be the probability of error when the
input to Alice is uniformly random but the input to Bob is~$i$. Note
that~$\epsilon = \sum_i \epsilon_i /n$. Let~$X$ denote the random
variable containing Alice's input, and let $M_B$ denote the qubits
held by Bob after he has received Alice's message, including his part
of the shared entangled state. From Properties~(\ref{eqn:infeq})
and~(\ref{eqn:infineq}) of mutual information in
Section~\ref{sec:info-background:entropy}, and the concavity of binary
entropy,
$$
I(X : M_B) \ge \sum_i I(X_i : M_B) \ge
\sum_i (1-H(\epsilon_i)) \ge n (1 - H(\epsilon)).
$$

The second inequality follows from the fact that Bob has a
measurement that predicts~$X_i$ with error~$\epsilon_i$ and
Fact~\ref{cl:mut} (Fano's inequality). On the other
hand,~$I(X:M_B)$ is bounded above by twice the number of qubits in
the message~\cite[Theorem~2]{CDNT98}. The lemma follows.
\end{proof}

Note that for public-coin randomized protocols we do not have the
factor of~$\frac{1}{2}$, and obtain a lower bound of~$n (1 -
H(\epsilon))$.

Next, we show how an efficient protocol for~$S_k$ gives rise to an
efficient protocol for~$S_{k-1}$. The intuition behind the argument is
the same as in proofs for classical communication~\cite{MNSW95,KN97}.
However, we use entirely new techniques from quantum information
theory, as developed in Section~\ref{sec:information-theory-new} and
also get better bounds.

\begin{proof}(Of Lemma~\ref{lem:reduction}):
For concreteness, we assume that~$k$ is even, so that~$\bar{P}_k$ is
Bob.
Let~$\cp$ be a protocol that solves~$S_{k}$ with respect to the uniform
distribution~$U$ with error~$\epsilon$, $k$ messages starting with Bob.
Let the communication complexity of~$\cp$ be~$\ell = \ell_1 +
\bar{\ell}$ with $\ell_1$ being the length of the first message sent.

Given the protocol~$\cp$,
we devise a protocol~$\cp'$ for solving~$S_{k-1}$
with respect to the uniform distribution, but with {\em Alice\/}
starting, and with only~$k-1$ messages.
The intuition behind the protocol~$\cp'$ is the following. It
first tries to recreate, from some shared prior entanglement,
the state after the first message in the
run of~$\cp$ on a specially chosen~$S_k$ instance, and then
simulates the remaining~$k-1$ rounds of communication of the
protocol~$\cp$ on the recreated state. The instance of~$S_k$ is such
that the solution to that instance coincides with the solution to
the given~$S_{k-1}$ instance. We thus get a protocol for~$S_{k-1}$ with the
desired properties. The details follow.

We start by describing the joint pure state that Alice and Bob
share in $\cp'$ prior to being given the inputs to the
problem~$S_{k-1}$. Consider the protocol~$\cp$ computing~$S_k$.
Let $M_A,M_B$ be the private qubits (or ``registers'') held by
Alice and Bob respectively. Let $Y = Y_1 Y_2 \cdots Y_n$ denote
the register containing the input to Bob. Consider the
state~$\ket{\chi}$ of the registers~$M_A M_B Y$, after Bob sends
the first message in~$\cp$, when~$Y$ is initialized to a uniform
superposition over~$\Y_k = \Y_{k-1}^n$. The prior entanglement
that Alice and Bob share in~$\cp'$ is then defined as

$$
    \frac{1}{\sqrt{n}} \; \sum_{j = 1}^n \;
    \ket{j}_A \ket{\chi}_{AB} \ket{j}_B,
$$
where the qubits~$M_A$ in~$\ket{\chi}$ are given to Alice
and~$M_B, Y$ to Bob. It simplifies the description of the protocol
if Alice and Bob measure the first and the last register,
respectively, of the shared state to get a common random index~$j
\in [n]$. Since these registers will not be modified during the
course of the protocol,  the behavior of~$\cp'$ is not affected by
this measurement.

We are ready to describe the steps of the protocol~$\cp'$. Given
the inputs~$x,y$ to~$S_{k-1}$,
\begin{enumerate}
\item
Alice, who gets the input $x$, initializes a register~$X$ to~$\ket{\phi}^{\tensor (j-1)}
\ket{x} \ket{\phi}^{\tensor (n-j)} \ket{j}$,
where~$\ket{\phi}$ is the uniform superposition over~$\X_{k-1}$.

Note that the state of the registers~$X M_A M_B Y$
is now exactly as after the first
message in a run of the protocol~$\cp$ on an input for~$S_k$ where~$a =
j$, all input registers~$X_i$ but for~$X_j$ are in uniform superposition
over~$\X_{k-1}$, $X_j = x$, and all~$Y_i$ are in uniform superposition
over~$\Y_{k-1}$.

\item
Bob, who gets the input $y$, applies a unitary
transformation~$V_{j,y}$ (to be defined below) to the
registers~$M_B Y$. This step is intended to bring the state of the
registers~$M_A M_B Y$ close to~$\ket{\chi(y)}$, the state after
the first message in a run of the protocol~$\cp$ on an input
for~$S_k$ with~$X, Y_1, Y_2, \ldots, Y_{j-1}, Y_{j+1}, \ldots,
Y_n$ as above, except that register~$Y_j$ is set to~$y$ rather
than the uniform superposition over~$\Y_{k-1}$. Note that on an
input as in~$\ket{\chi(y)}$, the result of a protocol for~$S_k$ is
expected to be the same as~$S_{k-1}(x,y)$.

\item
Alice and Bob now simulate the protocol~$\cp$ from the second message
onwards starting with the registers~$X M_A M_B Y$, and declare the
result of that procedure as the output of the protocol~$\cp'$.
\end{enumerate}

The transformation~$V_{j,y}$ is defined as follows. Consider the
state~$\ket{\chi(j,y)}$ of the registers~$M_A M_B Y$ (analogous
to~$\ket{\chi}$) obtained by running~$\cp$ till the first message
is sent, when the register~$Y$ is initialized
to~$\ket{\psi}^{\tensor (j-1)} \ket{y} \ket{\psi}^{\tensor
(n-j)}$, where~$\ket{\psi}$ is the uniform superposition
over~$\Y_{k-1}$. Let~$\rho = \trace_{M_B Y} \density{\chi}$,
and~$\rho_{j,y} = \trace_{M_B Y} \density{\chi(j,y)}$ be the
restriction of the two states to Alice. The
transformation~$V_{j,y}$ is defined as the local unitary operator
on~$M_B Y$, given by Theorem~\ref{thm:jozsa}, that achieves the
fidelity between~$\rho$ and~$\rho_{j,y}$. This completes the
description of~$\cp'$.

Observe that~$\cp'$ has~$k-1$ messages starting with Alice, and has
complexity~$\bar{\ell}$. We now analyze its probability of
error, under a uniform distribution on inputs.

Bob's part of the input
to~$S_k$ in~$\ket{\chi}$ and~$\ket{\chi(j,y)}$ differ only in the
register~$Y_j$: in the first state, this is uniform over~$\Y_{k-1}$,
whereas in the second state, this is set to~$y$. Thus,
the state~$\ket{\chi}$ when restricted to Alice
is the {\em average encoding}, over all~$y \in \Y_{k-1}$, of the
state~$\ket{\chi(j,y)}$ restricted to her:
$$
\rho ~~=~~ \frac{1}{\size{\Y_{k-1}}} \sum_{z \in \Y_{k-1}} \rho_{j,z}.
$$
The Average encoding theorem tells us that~$\rho$ and~$\rho_{j,y}$ are
close to each other on average, provided the mutual information~$\mu_j =
I(Y_j : M_A)$ between Alice's state and~$Y_j$ in a run of~$\cp$
on the uniform distribution on all inputs is small:

\begin{equation}
\label{eqn-err1}
\frac{1}{\size{\Y_{k-1}}} \sum_z \hels{\rho, \rho_{j,z}}
    ~~\le~~ \left( \frac{\ln 2}{2} \right) \mu_j.
\end{equation}

As in the proof of Lemma~\ref{lem:k=1}, it is not hard to see that if
the length~$\ell_1$ of the first message~$M$ is small relative to~$n$,
then for a random~$j$, this mutual information is small.
\begin{claim}
\label{cl:j}
$\sum_i \mu_i \le 2 \ell_1$. Thus,
$\E_j \; \mu_j \le 2\ell_1/n$.
\end{claim}

By Lemma \ref{lem:close}, the transformation~$V_{j,y}$
maps~$\ket{\chi}$ to a state close to~$\ket{\chi(j,y)}$, and by
Lemma~\ref{lem:tracevshellinger}
\begin{eqnarray}
\lefteqn{\trnorm{ \density{V_{j,y}\chi} - \density{\chi(j,y)}}} \nonumber \\
    & \le & 2\sqrt{2} \;\; \hel{ \density{V_{j,y}\chi}, \density{\chi(j,y)}
                             } \nonumber \\
    & = & 2\sqrt{2} \;\; \hel{ \rho,\rho_{j,y}}.  \label{eqn-err2}
\end{eqnarray}
For a random~$y \in \Y_{k-1}$, and a random~$j \in [n]$, then, the
average error in approximating the state~$\ket{\chi(j,y)}$ is
\begin{align*}
\lefteqn{\E_{j,y} \trnorm{ \density{V_{j,y}\chi} - \density{\chi(j,y)}}} \\
    & \le~~  2\sqrt{2} \;\; \E_{j,y} \, \hel{\rho, \rho_{j,y}} 
          &  {\textrm{From equation~(\ref{eqn-err2})}}  \\
    & \le~~ 2\sqrt{2}\;\; \E_j \left[ \E_y \, \hels{\rho, \rho_{j,y}}
                   \right]^{1/2} & {\textrm{By Jensen's inequality}}
    \\
    & \le~~ 2\sqrt{2}\;\; \E_j \left( \frac{\ln 2}{2} \mu_j
    \right)^{1/2} & {\textrm{From equation~(\ref{eqn-err1})}} \\
    & \le~~ 2\sqrt{\ln 2}\;\; \left[ \E_j \mu_j \right]^{1/2} 
            & {\textrm{~~By Jensen's inequality}} \\
    & \le~~ 3\, (\ell_1/n)^{1/2}. & {\textrm{~~From Claim~\ref{cl:j}}}
\end{align*}

Running the protocol~$\cp$ on the input described in step~2
of~$\cp'$ finds~$S_{k-1}(x,y)$ with probability of error at
most~$\epsilon$ on average when~$x,y$ are chosen at random.
Thus, running the protocol~$\cp$ on the state resulting from step~2 of the
protocol~$\cp'$ gives us the answer to~$S_{k-1}(x,y)$ with
average probability of error
only slightly higher than~$\epsilon$:
$$
\epsilon'
    ~~=~~ \epsilon + \frac{1}{2}\; \E_{j,y} \trnorm{ \density{V_{j,y}\chi}
                                   -  \density{\chi(j,y)}}
    ~~\le~~ \epsilon + 2\, (\ell_1/n)^{1/2},
$$
as claimed.
\end{proof}

For classical randomized protocols, it is possible to simplify the
reduction of~$S_{k-1}$ to~$S_k$ described above: This is accomplished
as follows. Recall that Alice and Bob share public random coins. They
use this to sample a (common) message~$m$ from the distribution over
classical messages in the first round of the protocol~$\cp$ for~$S_k$,
where the inputs are chosen uniformly at random.  They also pick a
common random index~$j \in [n]$. Alice now picks~$X_i$, $i\not=j$
uniformly at random from~$\X_{k-1}$, and sets~$X_j = x$, and~$a =
j$. Bob picks~$Y_1,\ldots,Y_n$ from the uniform distribution
over~$\Y_{k-1}^{j-1} \times \set{y} \times \Y_{k-1}^{n-j}$,
conditioned on the first message in the protocol~$\cp$ on such a
random input being equal to~$m$.  The distance between the joint state
so constructed and the joint state in the original protocol differs
(in~$\ell_1$-distance) by at most the distance between Alice's
marginal distributions.  Alice and Bob now simulate the protocol~$\cp$
from the second message onwards on the input~$X,Y$. A straightforward
analysis using the Average encoding theorem shows that the initial
state (consisting of the message and the inputs) constructed above
differs from the corresponding state in the protocol~$\cp$ by
only~$(2\ell_1/n)^{1/2}$. This simpler argument was noted in
Ref.~\cite{Maneva01} and independently in Ref.~\cite{Sen03}.

\subsection{The Disjointness Problem}\label{sec:disj}

We now investigate the bounded round complexity of the disjointness
problem. Here Alice and Bob each receive the incidence vector of a
subset of a size $n$ universe. They reject iff the sets are
disjoint. It is known~\cite{Klauck00,BW01} that
$Q_\epsilon^{1}(\disj)\ge(1-H(\epsilon))n$ and
$\eqcc_\epsilon^1(\disj)\ge(1-H(\epsilon))n/2$.  Furthermore
$Q^{O(\sqrt{n})}_{1/3}(\disj)=O(\sqrt{n}\log n)$ by an application of
Grover search~\cite{bcw}. This upper bound was later
improved~\cite{AA05} to~$O(\sqrt{n}\,)$, although the number of rounds
remained~$O(\sqrt{n}\,)$. We now prove a lower bound by reduction.

\begin{proof}(Of Corollary~\ref{thm-disj}):
Suppose we are given a $k$ round quantum protocol for the
disjointness problem having error $1/3$ and using~$c$ qubits.
W.l.o.g.\ we can assume Bob starts the communication, because the
problem is symmetrical, and that $k$ is even. We reduce the
communication problem $S_k$ from Section~\ref{sec:problem} to
\disj.

We visualize an instance of $S_k$ as defining a subtree of the
$n$-ary tree with $k+1$ levels and the edges at alternate levels
known to Alice and Bob, respectively. The leaves of the tree are
labelled by Boolean values known to Alice (since $k$ is even). The
only edge at the root connects it to the~$a$'th child, where~$a \in
[n]$ is the input that specifies which instance of~$S_{k-1}$ is to
be solved. The subtrees at the second level are defined
recursively according to the~$n$ instances of~$S_{k-1}$.

There are at most $n^k$ possible paths of length $k$ that could
start at the root vertex. With each such path we associate an
element in the universe for the disjointness problem. Given the
edges originating from each of their levels, Alice and Bob
construct an instance of $\disj$ on a universe of size $N = n^k$.
Alice checks for each possible path of length $k$ whether the path
is consistent with her input and whether the paths lead to a leaf
which corresponds to the bit 1. In this case she takes the
corresponding element of the universe into her subset. Bob
similarly constructs his subset. Now, if the two subsets
intersect, then the (unique) element in the intersection witnesses
a length $k$ path leading to 1-leaf. If the subsets do not
intersect, then the length $k$ path from the root leads to a
0-leaf.

We thus obtain a $k$ round protocol for $S_k$ in which Bob starts.
By Lemma~\ref{lem:lbi}, the communication~$c$ is $\Omega(n)$ for
any constant $k$. Since the input length for the constructed
instance of \disj\ is $N=n^k$, we get
$\eqcc^k_{1/3}(\disj)=\Omega(N^{1/k})$ for $k=O(1)$.
\end{proof}

\subsection{Beyond A Constant Number Of Messages}
\label{sec:non-const}

So far, we have discussed the complexity of solving~$S_k$ in the
context of protocols with a constant number of messages. In fact,
we may derive a meaningful lower bound even when~$k$ grows as a
function of the parameter~$n$ (hence as a function of~$N = n^k$,
the input length). We may state the result as follows.

\begin{theorem}
\label{thm:nonconst-rounds} For all~$k = k(n) \ge 1$ and
constant~$\epsilon < \frac{1}{2}$ we have $
\eqcc^{k,\bar{P}_k}_{U,\epsilon}(S_k)
    \;\;=\;\; \Omega\left( \frac{n}{k} + k \right).
$
\end{theorem}

\begin{proof}
Let $\ell = \eqcc^{k,\bar{P}_k}_{U,\epsilon}(S_{k})$. Then, there is a
protocol that achieves this communication complexity
with~$\ell_1,\ell_2,\ldots,\ell_k$ qubits of communication in the~$k$
rounds, respectively.
By repeated application of
Lemma~\ref{lem:reduction} there is a quantum protocol that solves
$S_1$ with one message, the wrong player starting, $\ell_k$
communication qubits and error
\begin{align*}
\epsilon_1 & ~~=~~ \epsilon+ 2 \sum_{i = 1}^{k-1} \left( \frac{\ell_i}{n}
                 \right)^{1/2} & \\
           & ~~\le~~ \epsilon+ 2 \left( \frac{k \sum_{i < k} \ell_i}{n}
                 \right)^{1/2} & \text{By Jensen's inequality} \\
           & ~~\le~~ \epsilon + 2 \left( \frac{k \ell}{n}
                 \right)^{1/2}
\end{align*}
For a constant $\delta \in (\epsilon,{1 \over 2})$, if $\ell \le
({\delta -\epsilon \over 2})^2\; {n \over k}$ then $\epsilon_1 \le
\delta$ and by Lemma~\ref{lem:k=1} we have $\ell \ge \ell_k \ge {{1 -
H(\delta)} \over 2} n$. This implies that $k \le ({\delta -\epsilon
\over 2})^2 \cdot 2 \cdot {1 \over 1-H(\delta)}$. For some $\delta$
close enough to $\epsilon$ we get $k<1$. A contradiction. This proves
that $\ell \ge \Omega({n \over k})$. Also, every $k$ round protocol
has at least $k$ communication qubits and so $\ell \ge k$.
\end{proof}
Note that this lower bound of~$\Omega(n/k + k)$ also applies to
classical randomized protocols.

The above theorem implies a gap in communication complexity
between~$k$ and~$k+1$ message protocols for~$k$ up
to~$\Theta((n/\log n)^{1/2}) = \Theta(\log N/\log\log N)$, and also
lower bounds for \disj\ for such~$k$.

\section{The Pointer Jumping Function}
\label{sec:pointer-jumping}

The pointer jumping function is considered in most results showing a
round-hierarchy for classical communication
complexity~\cite{DGS87,NW93,PRV99,Kl98}. This problem is a
particularly natural candidate for such results.

\begin{definition}[Pointer Jumping]
Let $V_A$ and $V_B$ be disjoint sets of $n$ vertices each.
Let $\F_A=\{f_A|f_A:V_A\to V_B\}$, and $\F_B=\{f_B|f_B:V_B\to V_A\}$,
and
$$
f(v)=f_{f_A,f_B}(v)
    =  \left\{
           \begin{array}{ll}
                f_A(v) & \textrm{ if } v\in V_A,\\
                f_B(v) & \textrm{ if } v\in V_B.
           \end{array}
       \right.
$$
Define $f^{(0)}(v)=v$ and $f^{(k)}(v)=f(f^{(k-1)}(v))$.

Then $g_k: \F_A\times \F_B\to (V_A\cup V_B)$ is defined by
$g_k(f_A,f_B)=f_{f_A,f_B}^{(k+1)}(v_1)$, where $v_1\in V_A$ is
fixed. The {\em pointer jumping function\/} $f_k: \F_A\times
\F_B\to\{0,1\}$ is the XOR of all the bits in the output of $g_k$.
\end{definition}
In the corresponding communication problem, Alice is given a
function~$f_A \in \F_A$, and Bob a function~$f_B \in \F_B$, and they are
required to compute~$f_k(f_A,f_B)$.

\subsection{Previous Work}

If Alice starts, $f_k$ has a deterministic $k$ round communication
complexity of $k\log n$. If Bob starts, Nisan and
Wigderson~\cite{NW93} proved that $f_k$ has a randomized $k$ round
communication complexity of $\Omega({n \over k^2}-k\log n)$. The lower
bound can also be improved to $\Omega({n \over k}+k)$, see
Klauck~\cite{Klauck00}. With techniques similar to the ones in this
section it is also possible to show a lower bound of
$\frac{(1-2\epsilon)^2n}{2k^2}-k\log n$ for the randomized $k$ round
complexity of $f_k$ when Bob starts. We omit the details.

The lower bounds are not far from the known upper bound. Nisan and
Wigderson~\cite{NW93} describe a randomized protocol for computing
$g_k$ with complexity $O({n\over k}\log n+k\log n)$ in the situation
where Bob starts and $k$ rounds are allowed. Ponzio
\etal{}~\cite{PRV99} show that when $k=O(1)$, the deterministic
communication complexity of $f_k$ is $O(n)$.

\subsection{A New Upper Bound}
\label{sec:pj-upper}

We first give a new classical upper bound which combines ideas from
Nisan and Wigderson~\cite{NW93} and Ponzio \etal{}~\cite{PRV99}. 
For~$n \ge 1$, define~$\log^{(1)}(n)=\log n$ and for~$k>1$, define
$$ \log^{(k)}(n) ~~=~~ \log (\max\{\log^{(k-1)}(n), 1\}). $$
Furthermore let
$\log^*(n)=\min\{k:\log^{(k)}(n)\le 1\}$.

\begin{theorem}\label{thm:rand-protocol}
$R_\epsilon^{k,B}(g_k) \le
O(k \log n ~+~ \frac{n}{k}\cdot \log{1 \over \epsilon} \cdot (\log^{(\lceil k/2\rceil)}(n) + \log k))$.
\end{theorem}

\begin{proof} The claim is trivial for $k=1$.

For greater $k$ Bob starts and we have the following protocol. At
the first round Bob guesses (with public random bits) a set $S_0$
of $\delta n$ random vertices from $V_B$, we specify $\delta$
later. For each chosen vertex $v$ Bob communicates the first
$\ell_0$ bits of $f_B(v)$, we specify $\ell_0$ later. Note that
the names of the chosen vertices are accessible to Alice without
communication, by reading the public random bits. The protocol
then proceeds in two stages.

\begin{itemize}
\item
Denote $v_t=f^{(t-1)}(v_1)$. For each round $i=1,\ldots,k$ the
active player sends $v_i$. I.e., at the first round Bob sends
nothing (as $v_1$ is known), at the second round Alice sends
$v_2=f(v_1)$, then Bob sends $f(v_2)$ and so on. Also, at each
round $i$ Alice checks whether $v_i \in S_0$. Let $t$ be the first
round in which this happens. If $t>{k \over 2}$ the two players
abort the protocol.

\item
The rounds $t,t+1,\ldots,k$ take a special form. Let us start with
round $t$. Alice knows $v_t \in S_0$ and therefore knows the first
$\ell_0$ bits of $f_B(v_t)$. Alice defines a set $S_1$ that
contains all elements of $V_A$ with that prefix. I.e.,  $|S_1| \le
{n \over 2^{\ell_0}}$ and $v_{t+1} = f(v_t) \in S_1$. For each $v
\in S_1$ Alice sends the first $\ell_1$ bits of $f_A(v)$. In
general, in the $(t+i)$'th round the active player knows
$\ell_{i}$ bits of $f(v_{t+i})$. The active player then defines a
set $S_{i+1}$ that contains all the elements of his side with that
prefix.  I.e.,  $|S_{i+1}| \le {n \over 2^{\ell_i}}$ and
$v_{t+i+1} = f(v_{t+i}) \in S_{i+1}$. For each $v \in S_{i+1}$ the
active player sends the first $\ell_{i+1}$ bits of $f(v)$.
\end{itemize}

We now specify the parameters. First we choose $\delta={4 \over k}
\ln {1 \over \epsilon}$. W.l.o.g. we can assume the vertices
$v_2,v_4,\ldots$ are all distinct, or Alice can easily save two
rounds and the players finish on time. For any choice of ${k \over
4}$ distinct vertices $v_2,\ldots,v_{k/2}$ the probability, over
the choice of $S_0$, that during the first ${k \over 2}$ rounds
Alice will not visit $S_0$ is at most $(1-{k \over 4n})^{\delta n}
\le e^{-{\delta k \over 4}} \le \epsilon$. So assume indeed that
$t \le {k \over 2}$.

We now chose $\ell_i=\log^{(\lceil k/2 \rceil - i)}n +3\log k$.
It follows that for some $i<{k \over 2}$
we have $\ell_i \ge \log n$ and $|S_i|=1$
and the active player who holds $v_{t+i}$ also knows
$f(v_{t+i})$, so he can save two
rounds and the computation ends on time.

We now count the number of communication bits. We need $k \log n$ bits
for communicating
$v_i$, $i=1,\ldots,k$. Also, we need $\sum_{i=0}^{\lceil k/2\rceil} |S_i| \ell_i$ bits
for communicating the first
$\ell_i$ bits of each element in $S_i$. Notice, however,
%
%
that $\ell_i \le {2^{\ell_{i-1}} \over k^2}$ and so:
\begin{eqnarray*}
\sum_{i=0}^{\lceil k/2\rceil}  |S_i| \ell_i
    & \le & n [ \delta \ell_0 + \sum_{i=1}^{\lceil k/2\rceil} {\ell_i \over 2^{\ell_{i-1}}}]
    ~ \le ~ n [ \delta \ell_0 + {1 \over k^2} \sum_{i = 1}^{\lceil k/2\rceil} 1 ] \\
    & = &    O({n \over k} \cdot \log{1 \over \epsilon}
             \cdot (\log^{(\lceil k/2 \rceil)}n + \log k))
\end{eqnarray*}
which completes the proof.
\end{proof}

\begin{corollary} If $k\ge 2\log^*(n)$ then $R_{1/3}^{k,B}(g_k)\le
  O((\frac{n}{k}+k) \log k)$.
\end{corollary}

\subsection{A Lower Bound On The Quantum Communication Complexity}
\label{sec:pj-lower}

In this section we prove a lower bound on the quantum
communication complexity of the pointer jumping function $f_k$,
for the situation that $k$ rounds are allowed and Bob sends the
first message. The proof uses the same ingredients as the proof of
the lower bound for the function $S_k$ in Theorem~\ref{thm:lb},
namely the Average Encoding Theorem and the Local Transition
Lemma. We will consider a quantity $d_t$ capturing the information
the active player has in round $t$ on vertex $t+1$ of the path.
This quantity will be the informational distance between the
active player's qubits and vertex $t+1$. Our goal will be to bound
$d_t$ in terms of $d_{t-1}$ (which is the information gain so far)
plus a term related to the average information on pointers in the
other player's input (which is low as long as the number of qubits
sent is small). This leads to a recursion imposing a lower bound
on the communication complexity, since in the end the protocol
must have reasonably large information to produce the output, and
in the beginning the corresponding information~$d_0$ is 0.

Let Alice be active in the $(t+1)$'th round. The informational distance
$d_{t+1}$ measures the distance between the state of, say, Alice's
qubits together with the next vertex  $F_B(V_{t+1})$ of the path,
and the tensor product of the states of Alice's qubits and
$F_B(V_{t+1})$. In the product state Alice has no information
about $F_B(V_{t+1})$, so if the two states are close Alice's
powers to say something about the vertex are very limited. We will
use the triangle inequality to bound $d_{t+1}$ by the sum of three
intermediate distances. In the first step we move from the state
given by the protocol to a state in which the $(t+1)$'th vertex is
replaced by a uniformly random vertex, independent of previous
communications. The penalty we have to pay for that is
proportional to $d_t$ which is a bound on the amount of
information Bob gained on $V_{t+1}$. We use the local transition
lemma to conceal Bob's ability to detect such a replacement. Once
the $(t+1)$'th vertex is random, we deal with the average
information a player (Bob) can get on a random pointer in the
other player's input, and this term is small when the number of
communicated qubits is small. The last step is similar to the
first and reverses the first one's effect, i.e., replaces the
``randomized'' $(t+1)$-th vertex by its real value again. We arrive
at the desired product state.

\begin{theorem}
\label{thm:pj-lb} $\eqcc^{k,B}_{1/8 }(f_k)\ge {n \over
{2^{O(k)}}}-k\log n$.
\end{theorem}
Note that the lower bound is linear in $n$ for constant $k$ and leads to
Theorem~\ref{thm:PJ-result}. It implies a separation between the~$k$
and~$k+1$ round complexity of Pointer Jumping for~$k$ upto~$\Theta(\log
n) = \Theta(\log N)$, where~$N = n\log n$ is the input size.

\begin{proof}(of Theorem~\ref{thm:pj-lb})
Fix a quantum protocol
for $f_k$ with probability of error~${1 \over 8}$, $k$ rounds, and with 
Bob starting.
Usually a protocol gets some classical $f_A$ and $f_B$ as inputs,
but we will investigate what happens if the protocol is started on
a superposition over all inputs, in which all inputs have the same
amplitude, i.e., on

$$
\sum_{f_A \in {\F_A}, {f}_B \in {\F_B}} \frac{1}{n^n} \ket{f_A}
\ket{ f_B}. $$

 Note that $|{\F}_A|=|{\F}_B| = {n}^{n}$. The
superposition over all inputs is measured after the protocol has
finished, so that a uniformly random input and the result of the
protocol on that input are produced.

We also require that before round $t$ the active player computes and measures the vertex
$v_t=f^{(t-1)}(v_1)$, and includes it in the message that is sent to the other player,
who stores it in some qubits $V_t$.
Thus, at the first round Bob sends $v_0$ (which is known in advance) to Alice,
at the second round Alice sends $v_2=F_A(v_1)$ to Bob and so on.
This increases the communication by an additive $k\log n$ term.
Notice that $F_A,F_B$ are in a uniform superposition over all possible inputs,
and so if we don't measure $F_A$ and $F_B$ the register
$V_i$ is also in a uniform superposition for every $i>1$.
The density matrix of the global state of the protocol before the communication of
round $t$ is
${\rho}_{M_{A,t}M_{B,t}F_AF_B}$, where $F_A,F_B$ are the qubits holding
the inputs of Alice and Bob and $M_{A,t}$ resp.~$M_{B,t}$ are the
other qubits in the possession of Alice and Bob before the
communication of round $t$. The state of the latter two systems of
qubits may be entangled. In the beginning these qubits are independent
of the input.
We also denote $\tilde{\rho}_{M_{A,t}M_{B,t}F_AF_B}$ the density matrix of the system in the
case where we do not measure any of the $V_i$.

Let us denote $d_{t}=D^2(M_{B,t}F_B:F_A(V_{t}))$ when $t$ is odd,
where the register~$F_A(V_{t})$ has been measured.
Notice that at this stage $V_t$ is measured and $F_A(V_t)$ is a subregister
of $F_A$.
The quantity~$d_t$ is a measure of Bob's information on
the value $F_A(v)$ Alice is going to compute.
We similarly let
$d_{t} = D^2(M_{A,t}F_A:F_B(V_{t}))$ when $t$  is even,
where the register~$F_B(V_t)$ has been measured.

We assume that the communication complexity of the protocol is
$\delta n$ and prove a lower bound $\delta \ge 2^{-{O(k)}}$.
The general strategy of the proof is induction over the rounds, to
successively bound~$d_1, d_2, \ldots, d_{k+1}$.
Bob sends the first message. As Bob has seen no message yet,
we have that $I(M_{B,1}F_B:F_A(V_1))=0$, and hence $d_1=0$.
We show that
\begin{lemma}
\label{lem:induction}
$d_{t+1} \le 8 d_t+4\delta$.
\end{lemma}

We see that
$d_{t+1} \le 9^t \delta$ for all $t \ge 0$.
After round $k$ one player, say Alice, announces the result
which is supposed to be the parity of $F_B(V_{k+1})$ and included in
$M_{A,k+1}$.
On the one hand $d_{k+1} ~=~ D^2(M_{A,k+1}:F_B(V_{k+1}))
\le 9^{k} \delta$.
On the other hand, by Lemma~\ref{lem:propID}(\ref{enu:BooleanX})
$D^2(M_{A,k+1}:\bigoplus F_B(V_{k+1}))\ge 1/8-1 / 16=1/16$.
Together,  ${1 \over 16} \le 9^k\delta$, so
$\delta \ge 2^{-{O(k)}}$.
\end{proof}

We now turn to proving Lemma \ref{lem:induction}.

\begin{proof}(Of Lemma \ref{lem:induction}):
W.l.o.g.\ let Alice be active in round $t+1$.
Let $M_A=M_{A,t+1}$ and $M_B=M_{B,t+1}$.
Before the $t+1$ round $V_{t+1}=F_A(V_{t})$ is measured.
The resulting state is a
probabilistic ensemble over the possibilities to fix
$V_1,\ldots,V_{t+1}$, which are then classically distributed.
Alice's reduced state is block
diagonal with respect to the possible values of the
vertices $V_1,\ldots,V_{t+1}$.
For any value $v$ of $V_{t+1}$ let
$\rho^{v}_{M_A M_B F_A F_B} = \rho^{V_{t+1} = v}_{M_A M_B F_A F_B}$
denote the pure state with vertex $V_{t+1}$ fixed to $v$.
Our first goal is to bound the amount of information Bob has at this stage about Alice's
value $V_{t+1}$. We define:

\begin{eqnarray*}
\gamma_{v} & \eqdef & \hels{\rho^{v}_{M_B F_B},\rho_{M_B F_B}}.
\end{eqnarray*}

I.e., we look at Bob's view before the $t+1$ message, and in
particular before Alice sends $V_{t+1}$ to him, and we let
$\gamma_{v}$ measure how much Bob's view when $V_{t+1}=v$ differs
from Bob's average view. We show that these two are typically
close to each other, namely:

\begin{lemma}
\label{lem:gamma}
$\E_{v} \gamma_{v}  \le  d_t$.
\end{lemma}
Loosely speaking this says that Bob does not know more than~$d_t$ units
of information about $F_A$.

The next step is to replace the actual state
$\rho^{v}_{M_AM_B{F_A}F_B}$ where $V_{t+1}=v$ with the average
case $\rho_{M_AM_B{F_A}F_BR}$ where nothing is known about
$V_{t+1}$. As we saw, typically, Bob can not distinguish between
the actual encoding and the average one, so this should not matter
much to Bob. We let  $\rho^{v}_{M_AM_B{F_A}F_BR}$ be a
purification of $\rho^{v}_{M_AM_B{F_A}F_B}$ where $R$ is some
additional space used to purify the random
path~$V_1,\ldots,V_{t}$. I.e., $\rho^{v}_{M_AM_B{F_A}F_BR}$
reflects a purification of Bob's view, when $V_{t+1}=v$. We let
$\rho_{M_AM_B{F_A}F_BR}$ be a purification of~
$\rho_{M_AM_B{F_A}F_B}$ where $R$ is some additional space used to
purify the random path~$V_1,\ldots,V_{t+1}$. Now, due to
Lemma~\ref{lem:close} there is a local unitary transformation
$U_v$ acting only on ${F_AM_AR}$ such that $
\sigma^{v}_{M_AM_B{F_A}F_BR}
    ~~\eqdef~~ U_v \rho_{M_AM_B{F_A}F_BR} U_v^\dagger,
$ and~$\rho^{v}_{M_AM_B{F_A}F_BR}$ are close to each other.
$\sigma^{v}_{M_AM_B{F_A}F_BR}$ reflects a purification of Bob's
average view with Alice locally adding $V_{t+1}=v$ to it . Notice
that in $\sigma^{v}_{M_AM_B{F_A}F_BR}$, $v$ is arbitrary and in
particular can be different than $V_{t+1}$. By Lemma
\ref{lem:close} for all vertices~$v \in V_B$,
\begin{eqnarray}
\hels{\rho^{v}_{M_A{F_A}}  , \sigma^{v}_{M_A{F_A}}}
   & \le & \hels{\rho^{v}_{M_A{F_A}F_B(v)}  , \sigma^{v}_{M_A{F_A}F_B(v)}}
           \nonumber \\
    & \le & \hels{\rho^{v}_{M_A{M_B}{F_A}F_BR}  ,
            \sigma^{v}_{M_A{M_B}{F_A}F_BR}} \nonumber \\
\label{eqn:rho-sigma}
    & = & \hels{\rho^{v}_{M_B F_B},\rho_{M_B F_B}} = \gamma_{v},
\end{eqnarray}

We are interested in the value
\begin{eqnarray*}
d_{t+1} ~~=~~ D^2(M_AF_A:F_B(V_{t+1}))
 &=& \E_{v} \;\; \hels{\rho^{v}_{M_AF_AF_B(v)}
              , \rho^{v}_{M_AF_A}\otimes\rho_{F_B(v)}},
\end{eqnarray*}
where $F_B(v)$ is measured and the expectation is over the uniform
distribution on vertices~$v$. We now study this expression under
the average case scenario, i.e., we look at $\hels{\sigma^{v}_{M_A
F_A F_B(v)} , \sigma^{v}_{M_A F_A} \otimes \rho_{F_B(v)}}$. We
prove:

\begin{lemma}
\label{lem:rho-sigma}
For all vertices~$v \in V_B$,
\begin{eqnarray*}
\hels{\sigma^{v}_{M_A F_A F_B(v)}
         , \sigma^{v}_{M_A F_A} \otimes \rho_{F_B(v)}}
    & \le &  \tilde{d}_{t+1}(v)
\end{eqnarray*}
\end{lemma}

where,
\begin{eqnarray}
\label{eqn:beta}
\tilde{d}_{t+1}(v) & \eqdef & \hels{\tilde{\rho}_{M_AF_AF_B(v)}
                           , \tilde{\rho}_{M_AF_A}
                             \otimes \rho_{F_B(v)}},
\end{eqnarray}

where $F_B(v)$ is assumed to have been measured. Recall that in
$\tilde{\rho}$ we let $V_1,\ldots,V_{t+1}$ go unmeasured and that
$v$ is an arbitrary value not necessarily equal to~$V_{t+1}$. We then
prove:
\begin{lemma}
\label{lem:tilded}
$\E_{v} ~\tilde{d}_{t+1}(v)  \le   2\delta$.
\end{lemma}

Assuming the above lemma, we see that for all~$v$:
\begin{align*}
\lefteqn{ \hel{ \rho_{M_AF_AF_B(v)}^{v}
          , \rho_{M_AF_A}^{v} \otimes \rho_{F_B(v)} }} \\
  & \le~~ \hel{ \rho_{M_AF_AF_B(v)}^{v} , \sigma^{v}_{M_AF_AF_B(v)}}
          & \\
  & ~~~ + \hel{ \sigma^{v}_{M_AF_AF_B(v)} ,
          \sigma^{v}_{M_AF_A} \otimes \rho_{F_B(v)} } & \\
  & ~~~ + \hel{ \sigma^{v}_{M_AF_A} \otimes \rho_{F_B(v)} ,
          \rho_{M_AF_A}^{v} \otimes \rho_{F_B(v)} }
          &  \\ 
  & \le~~  2\sqrt{\gamma_{v}}
           + \hel{ \sigma^{v}_{M_AF_AF_B(v)}
                   , \sigma^{v}_{M_AF_A} \otimes \rho_{F_B(v)}} &
           \textrm{From equation~(\ref{eqn:rho-sigma})} \\
   & \le~~  2\sqrt{\gamma_{v}}+\sqrt{\tilde{d}_{t+1}(v)} &
           \textrm{From Lemma~(\ref{lem:rho-sigma})}.
\end{align*}
Squaring both sides,
\begin{align}
\hels{ \rho_{M_AF_AF_B(v)}^{v}
          , \rho_{M_AF_A}^{v} \otimes \rho_{F_B(v)} }
& ~~\le~~  \left(2\sqrt{\gamma_{v}}+\sqrt{\tilde{d}_{t+1}(v)}\right)^2
        \notag \\
\label{eqn-dist}
& ~~\le~~  8 \gamma_v+2\tilde{d}_{t+1}(v).
\end{align}
I.e., we paid an~$8\gamma_v$ penalty, and we switched to the scenario where
Bob has no information about $V_{t+1}$. Now,
\begin{align*}
 D^2(M_AF_A:F_B(V_{t+1}))
 & ~~=~~   \E_{v}\;\; \hels{ \rho_{M_AF_AF_B(v)}^{v},
         \rho_{M_AF_A}^{v} \otimes \rho_{F_B(v) } } & \\
 & ~~\le~~ \E_{v }\; [8\gamma_{v }+2\tilde{d}_{t+1}(v)] & \textrm{By
         equation~(\ref{eqn-dist})} \\
 & ~~\le~~ 8d_t+4\delta & \textrm{By Lemma~\ref{lem:tilded}}.
\end{align*}
This completes the proof of Lemma~\ref{lem:induction}.
\end{proof}

We finish the proof of Theorem~\ref{thm:pj-lb} by proving the
remaining Lemmas.

\begin{proof}(Of Lemma \ref{lem:gamma}):
By definition
$\E_{v} \gamma_{v}$
is $\E_{u} \hels{ \rho^{V_t=u}_{M_BF_B F_A(u)} , \rho^{V_t=u}_{M_BF_B}
                                 \otimes \rho_{F_A(u)} }
    =  D^2(M_B F_B : F_A(V_t))$.
Now, $D^2(M_{B,t+1} F_B : F_A(V_t)) \le D^2(M_{B,t} F_B : F_A(V_t)) = d_t$
because Bob sends the $t$'th message,
and this only decreases the informational distance.
\end{proof}

\begin{proof}(Of Lemma \ref{lem:rho-sigma}):

\begin{align}
\lefteqn{ \hels{ \sigma^{v}_{M_A{F_A}F_B(v)}
                , \sigma^{v}_{M_A{F_A}} \otimes \rho_{F_B(v)}} } \notag \\
  & \le~~ \hels{ \sigma^{v}_{M_A{F_A}RF_B(v)}
              , \sigma^{v}_{M_A{F_A}R} \otimes \rho_{F_B(v)}}
        &  \notag \\
  & =~~ \hels{ \rho_{M_A{F_A}RF_B(v)}
              , \rho_{M_A{F_A}R} \otimes \rho_{F_B(v)} }
        & \textrm{By unitarity} \notag \\
  & =~~ \hels{ \tilde{\rho}_{M_A{F_A}F_B(v)}
              , \tilde{\rho}_{M_A{F_A}} \otimes \rho_{F_B(v)} }
        &  \label{eqn-star} \\
  & =~~ \tilde{d}_{t+1}(v) & \textrm{By definition (\ref{eqn:beta})}.
\notag
\end{align}

For equation~(\ref{eqn-star}),
notice that $R$ holds the path~$V_1,\ldots,V_{t+1}$,
which is determined by~$M_A F_A$.
We can apply a unitary transformation that ``erases''
this. We then get a pure state that is $\rho$ with $V_1,\ldots,V_{t+1}$ unmeasured,
i.e., what we called $\tilde{\rho}$
\end{proof}

\begin{proof}(Of Lemma \ref{lem:tilded}):
We first bound the information Alice has on Bob's input.
For all $t$, ~$I(M_{A,t}F_A:F_B)$ is bounded above by twice
the number of qubits in the messages so far due to
Lemma~\ref{lem:factor2}, assuming that $F_B$ is measured,  i.e.,
$I(M_{A,t}F_A:F_B) \le 2\delta n$.
%
%
Thus considering the situation that $F_B$ is distributed uniformly
instead of being in the uniform superposition
we get $\E_v I(M_A F_A:F_B(v)) \le 2\delta$ (where~$v$ is uniformly
random), using Equation (\ref{eqn:infeq}) and that the
$F_B(v)$ are mutually independent.
Now,
\begin{eqnarray*}
\E_{v} \tilde{d}_{t+1}(v)
    & = &  \E_v \hels{\tilde{\rho}_{M_AF_AF_B(v)}
                     , \tilde{\rho}_{M_AF_A} \otimes \rho_{F_B(v)}} \\
    & = &  \E_v D^2(M_{A} F_A:F_B(v)),
\end{eqnarray*}
where~$F_A M_A M_B F_B$ are as in the protocol without
measurements. Also $I(M_AF_A:F_B(v))$ is invariant if $F_B(i)$ is
in superposition or measured for $i\neq v$. So,

\begin{align*}
 \E_v D^2(M_{A} F_A:F_B(v)) & ~~\le~~ \E_v \, I(M_{A}F_A:F_B(v))
        & \textrm{~~By Lemma \ref{lem:inf-dist}} \\
  & ~~=~~  2\delta. &
\end{align*}
\end{proof}

\subsection*{Acknowledgements}

We thank Jaikumar Radhakrishnan and Venkatesh Srinivasan for their
input on the classical communication complexity of Pointer Jumping
and the subproblem~$S_k$, Dorit Aharonov and Pranab Sen for
helpful feedback on earlier versions of the paper, and Elitza
Maneva and Leonard Schulman for discussions on applying our
techniques to classical protocols for~$S_k$. We thank the
anonymous referee for useful comments.

\bibliographystyle{IEEEbib}
\bibliography{refs}

\begin{thebibliography}{10}

\bibitem{bcw}
H.~Buhrman, R.~Cleve, and A.~Wigderson,
\newblock ``Quantum vs.~classical communication and computation,''
\newblock in {\em Proceedings of the 30th Annual ACM Symposium on Theory of
  Computing}, 1998, pp. 63--68.

\bibitem{ASTVW03}
A.~Ambainis, L.~J. Schulman, A.~Ta-Shma, U.~Vazirani, and A.~Wigderson,
\newblock ``The quantum communication complexity of sampling,''
\newblock {\em SIAM Journal on Computing}, vol. 32, no. 6, pp. 1570--1585,
  2003.

\bibitem{R99}
R.~Raz,
\newblock ``Exponential separation of quantum and classical communication
  complexity,''
\newblock in {\em Proceedings of the 31st Annual ACM Symposium on Theory of
  Computing}, 1999, pp. 358--367.

\bibitem{KitaevW00}
A.~Kitaev and J.~Watrous,
\newblock ``Parallelization, amplification, and exponential time simulation of
  quantum interactive proof systems,''
\newblock in {\em Proceedings of the 32nd Annual ACM Symposium on Theory of
  Computing}, 2000, pp. 608--617.

\bibitem{PS82}
C.H. Papadimitriou and M.~Sipser,
\newblock ``Communication complexity,''
\newblock in {\em Proceedings of the 14th Annual ACM Symposium on Theory of
  Computing}, 1982, pp. 196--200.

\bibitem{DGS87}
P.~Duris, Z.~Galil, and G.~Schnitger,
\newblock ``Lower bounds on communication complexity,''
\newblock {\em Information and Computation}, vol. 73(1), pp. 1--22, 1987.

\bibitem{NW93}
N.~Nisan and A.~Wigderson,
\newblock ``Rounds in communication complexity revisited,''
\newblock {\em SIAM Journal on Computing}, vol. 22(1), pp. 211--219, 1993.

\bibitem{Kl98}
H.~Klauck,
\newblock ``Lower bounds for computation with limited nondeterminism,''
\newblock in {\em Proceedings of the 13th Annual IEEE Conference on
  Computational Complexity}, 1998, pp. 141--153.

\bibitem{PRV99}
S.J. Ponzio, J.~Radhakrishnan, and S.~Venkatesh,
\newblock ``The communication complexity of pointer chasing, applications of
  entropy and sampling,''
\newblock in {\em Proceedings of the 31st Annual ACM Symposium on Theory of
  Computing}, 1999, pp. 602--611.

\bibitem{MNSW95}
P.B. Miltersen, N.~Nisan, S.~Safra, and A.~Wigderson,
\newblock ``On data structures and asymmetric communication complexity,''
\newblock in {\em Proceedings of the 27th Annual ACM Symposium on Theory of
  Computing}, 1995, pp. 103--111.

\bibitem{KNTZ01}
H.~Klauck, A.~Nayak, A.~Ta-Shma, and D.~Zuckerman,
\newblock ``Interaction in quantum communication and the complexity of set
  disjointness,''
\newblock in {\em Proceedings of the 33rd Annual ACM Symposium on Theory of
  Computing}, 2001, pp. 124--133.

\bibitem{BW01}
H.~Buhrman and R.~de~Wolf,
\newblock ``Communication complexity lower bounds by polynomials,''
\newblock in {\em Proceedings of the 16th Annual IEEE Conference on
  Computational Complexity}, 2001.

\bibitem{Nayak99}
A.~Nayak,
\newblock ``Optimal lower bounds for quantum automata and random access
  codes,''
\newblock in {\em Proceedings of the 40th Annual IEEE Symposium on Foundations
  of Computer Science}, 1999, pp. 369--376.

\bibitem{AmbainisNTV02}
Andris Ambainis, Ashwin Nayak, Amnon Ta-Shma, and Umesh Vazirani,
\newblock ``Dense quantum coding and quantum finite automata,''
\newblock {\em Journal of the ACM}, vol. 49, no. 4, pp. 1--16, July 2002.

\bibitem{CDNT98}
R.~Cleve, W.~van Dam, M.~Nielsen, and A.~Tapp,
\newblock ``Quantum entanglement and the communication complexity of the inner
  product function,''
\newblock in {\em QCQS: NASA International Conference on Quantum Computing and
  Quantum Communications, QCQS}. 1998, LNCS.

\bibitem{BW92}
C.H. Bennett and S.J. Wiesner,
\newblock ``Communication via one- and two-particle operators on einstein-
  podolsky-rosen states,''
\newblock {\em Physical review letters}, vol. 69, pp. 2881--2884, 1992.

\bibitem{P98}
J.~Preskill,
\newblock ``Lecture notes.,''
  http:/$\!$/www.theory.caltech.edu/people/preskill/ph229/$\!$., 1998.

\bibitem{NC00}
M.A. Nielsen and I.L. Chuang,
\newblock {\em Quantum Computation and Quantum Information},
\newblock Cambridge University Press, Cambridge, 2000.

\bibitem{R03}
A.A. Razborov,
\newblock ``Quantum communication complexity of symmetric predicates,''
\newblock {\em Izvestiya of the Russian Academy of Science, Mathematics}, vol.
  67, pp. 145--159, 2003,
\newblock see also quant-ph/0204025.

\bibitem{AA05}
Scott Aaronson and Andris Ambainis,
\newblock ``Quantum search of spatial regions,''
\newblock {\em Theory of Computing}, vol. 1, pp. 47--79, 2005.

\bibitem{JRS03}
R.~Jain, J.~Radhakrishnan, and P.~Sen,
\newblock ``A lower bound for bounded round quantum communication complexity of
  set disjointness,''
\newblock in {\em Proceedings of the 44th Annual IEEE Symposium on Foundations
  of Computer Science}, 2003, pp. 220--229,
\newblock see also quant-ph/0303138.

\bibitem{JRS02a}
R.~Jain, J.~Radhakrishnan, and P.~Sen,
\newblock ``The quantum communication complexity of the pointer chasing
  problem: the bit version,''
\newblock in {\em Proceedings of the 22nd Conference on Foundations of Software
  Technology and Theoretical Computer Science}, 2002, pp. 218--229.

\bibitem{JRS02b}
R.~Jain, J.~Radhakrishnan, and P.~Sen,
\newblock ``Privacy and interaction in quantum communication complexity and a
  theorem about the relative entropy of quantum states,''
\newblock in {\em Proceedings of the 43rd Annual IEEE Symposium on Foundations
  of Computer Science}, 2002, pp. 429--438.

\bibitem{SV01}
S.~Venkatesh and P.~Sen,
\newblock ``Lower bounds in the quantum cell probe model,''
\newblock in {\em Proceedings of the 28th International Colloquium on Automata,
  Languages, and Programming}, 2001, pp. 358--369,
\newblock see also cs.CC/030903.

\bibitem{K04}
H.~Klauck,
\newblock ``Quantum and approximate privacy,''
\newblock {\em Theory of Computing Systems}, vol. 37, pp. 221--246, 2004.

\bibitem{CR04}
Amit Chakrabarti and Oded Regev,
\newblock ``An optimal randomised cell probe lower bound for approximate
  nearest neighbour searching,''
\newblock in {\em Proceedings of the 45th Annual IEEE Symposium on Foundations
  of Computer Science}, 2004.

\bibitem{Jozsa94}
R.~Jozsa,
\newblock ``Fidelity for mixed quantum states.,''
\newblock {\em Journal of Modern Optics}, vol. 41(12), pp. 2315--2323, 1994.

\bibitem{Uhlmann76}
A.~Uhlmann,
\newblock ``The `transition probability' in the state space of a
  $*$-algebra.,''
\newblock {\em Reports on Mathematical Physics}, vol. 9, pp. 273--279, 1976.

\bibitem{FG99}
C.A. Fuchs and J.~van~de Graaf,
\newblock ``Cryptographic distinguishability measures for quantum-mechanical
  states.,''
\newblock {\em IEEE Transactions on Information Theory}, vol. 45(4), pp.
  1216--1227, 1999.

\bibitem{LC98}
H.~Lo and H.~Chau,
\newblock ``Why quantum bit commitment and ideal quantum coin tossing are
  impossible.,''
\newblock {\em Physica D}, vol. 120, pp. 177--187, 1998,
\newblock see also quant-ph/9711065.

\bibitem{M97}
D.~Mayers,
\newblock ``Unconditionally secure quantum bit commitment is impossible,''
\newblock {\em Physical review letters}, vol. 78, pp. 3414--3417, 1997.

\bibitem{CT91}
T.~M. Cover and J.~A. Thomas,
\newblock {\em Elements of Information Theory},
\newblock Wiley Series in Telecommunications. John Wiley \& Sons, New York, NY,
  USA, 1991.

\bibitem{OP93}
Masanori Ohya and Denes Petz,
\newblock {\em Quantum Entropy and its Use},
\newblock Texts and Monographs in {P}hysics. Springer-Verlag, Heidelberg, 1993,
\newblock Second edition, 2004.

\bibitem{D78}
D.~Dacunha-Castelle,
\newblock ``Vitesse de convergence pour certains problemes statistiques,''
\newblock in {\em Ecole d'Ete de Probabilites de Saint-Flour VII-1977, Lecture
  Notes in Mathematics 678}, 1978, pp. 1--172.

\bibitem{Y93}
A.C.-C. Yao,
\newblock ``Quantum circuit complexity,''
\newblock in {\em Proceedings of the 34th Annual IEEE Symposium on Foundations
  of Computer Science}, 1993, pp. 352--361.

\bibitem{KN97}
E.~Kushilevitz and N.~Nisan,
\newblock {\em Communication Complexity},
\newblock Cambridge University Press, Cambridge, 1997.

\bibitem{Maneva01}
Elitza Maneva,
\newblock ``Interactive communication on noisy channels,'' B.S. Thesis,
  California Institute of Technology, Pasadena, CA, USA, 2001.

\bibitem{Sen03}
Pranab Sen,
\newblock ``Lower bounds for predecessor searching in the cell probe model,''
\newblock in {\em Proceedings of the 18th Annual IEEE Conference on
  Computational Complexity}, 2003, pp. 73--83.

\bibitem{Klauck00}
H.~Klauck,
\newblock ``On quantum and probabilistic communication: Las vegas and one-way
  protocols,''
\newblock in {\em Proceedings of the 32nd Annual ACM Symposium on Theory of
  Computing}, 2000, pp. 644--651.

\end{thebibliography}

\end{document}